\documentclass[journal,12pt,draftclsnofoot,onecolumn]{IEEEtran}
\IEEEoverridecommandlockouts
\usepackage{verbatim}
\usepackage{cite}
\usepackage{amsmath,amssymb,amsfonts}
\usepackage{graphicx}
\usepackage{textcomp}
\usepackage{xcolor}
\usepackage{epsfig}
\newtheorem{Remark}{\it Remark}[section]

\newtheorem{Proposition}{\it Proposition}[section]
\newtheorem{Lemma}{\it Lemma}[section]

\makeatletter
\newcommand{\Rmnum}[1]{\expandafter\@slowromancap\romannumeral #1@}
\makeatother
\usepackage{bm}
\usepackage{multicol}
\usepackage{setspace}
\usepackage{subfigure}
\usepackage{epstopdf}
\usepackage{fancyhdr} 
\usepackage{indentfirst}
\usepackage{enumerate}
\usepackage{stfloats}
\usepackage{bm}
\usepackage{multirow}
\usepackage{threeparttable}
\usepackage{CJK}
\usepackage[justification=centering]{caption}
\usepackage{makecell}
\usepackage{caption}
\usepackage{cases}
\usepackage{amsmath}
\usepackage{algorithm}
\usepackage{algpseudocode}

\usepackage{graphics}
\usepackage{caption}
\usepackage{booktabs}

\definecolor{deepblue}{rgb}{0.6,0,0.4}
\definecolor{red}{rgb}{0.8,0,0}
\definecolor{blue}{rgb}{0.2,0,0.8}

\def\BibTeX{{\rm B\kern-.05em{\sc i\kern-.025em b}\kern-.08em
		T\kern-.1667em\lower.7ex\hbox{E}\kern-.125emX}}

\begin{document}

 \title{ Design and Performance  of  Resonant Beam Communications---Part I: Quasi-Static Scenario}

 \author{\IEEEauthorblockN{Dongxu Li, Yuanming Tian, Chuan Huang, Qingwen Liu, and Shengli Zhou}
 \thanks{
Dongxu Li and Yuanming Tian are with the Future Network of Intelligence Institute and the School of Science and Engineering, the Chinese University of Hong Kong, Shenzhen 518172, China (emails: dongxuli@link.cuhk.edu.cn; yuanmingtian@link.cuhk.edu.cn).

Chuan Huang is with the School of Science and Engineering and the Future Network of Intelligence Institute, the Chinese University of Hong Kong, Shenzhen 518172, China (email: huangchuan@cuhk.edu.cn).

    Qingwen Liu is with the College of Electronics and Information Engineering, Tongji University, Shanghai 201804, China (email: qliu@tongji.edu.cn).

    Shengli Zhou is with the Department of Electrical and Computer Engineering, University of Connecticut, Storrs, CT 06250 USA (email: shengli.zhou@uconn.edu).
}
 }
 \maketitle
 
\begin{abstract}

  This two-part paper studies a point-to-point resonant beam communication (RBCom) system, where two separately deployed retroreflectors are adopted to generate the resonant beam between the transmitter and the receiver, and analyzes the transmission rate of the considered system under both the quasi-static and mobile scenarios. Part I of this paper focuses on the quasi-static scenario where the locations of the transmitter and the receiver are relatively fixed.  Specifically, we propose a new information-bearing scheme which adopts a synchronization-based amplitude modulation method to mitigate the echo interference caused by the reflected resonant beam.  With this scheme, we show that the quasi-static RBCom channel is equivalent to a Markov channel and can be further simplified as an amplitude-constrained additive white Gaussian noise channel.  Moreover, we develop an algorithm that jointly employs the bisection and exhaustive search to maximize its capacity upper and lower bounds. Finally, numerical results validate our analysis. Part II of this paper discusses the performance of the RBCom system under the mobile scenario.

	\end{abstract}
	\section{Introduction}
	Optical wireless communication (OWC) \cite{Elgala} is an emerging technology with the potential to enable extremely high data transmission rates in future 6G wireless systems \cite{Chowdhury}. Nearly 400 THz frequency band occupied by OWC provides sufficient communication resources to support the target Tbps data transmissions in 6G communications \cite{Karunatilaka}. Moreover, OWC has other significant advantages over the RF technologies such as robustness to electromagnetic interference \cite{Uysal} and no spectrum licensing requirements \cite{chowdhury2019role} for the implementations.   
	
	  Existing OWC systems can be classified into two categories based on their operating frequencies \cite{owc1,Pathak,vlc}: light-emitting-diode (LED)-based OWC which works in the visible spectrum  ($390$-$750$ nm),  and laser-based OWC which operates in the infrared band ($750$-$1600 $ nm). LED-based OWCs use an LED luminaire to produce light waves, whose intensity is adopted to carry information. The LED luminaire emits light at a quite wide angle, making the alignment between the transmitter and the receiver relatively simple. The authors in \cite{Klaver} designed an LED front end using $20$ equally spaced LEDs to provide $360^{\circ}$ coverage. However, the LED-based OWCs are mainly deployed for indoor scenarios, since the path loss from the transmitter to the receiver increases dramatically as the distance between them grows. The authors in \cite{y_Wang}  reported an experimental achievement of $8$ Gbps data rate using one single LED for indoor transmissions.  On the other hand, due to the monochromaticity, coherency, and collimation of lasers, laser-based OWCs can achieve very high data rate when the transmitter and the receiver are well aligned \cite{fso}. The authors in \cite{Tsonev} showed that laser-based OWCs are capable of attaining a transmission rate of $100$ Gbps under standard indoor illumination levels.   In \cite{Lange}, an OWC system transmitted data at a rate of $10$ Gbps over $10000$ km with a $100$ W laser power and a bit error rate of $10^{-9}$. However, in order to guarantee the link connectivity between the transmitter and the receiver, an acquisition, tracking, and pointing (ATP) subsystem is necessary to adaptively track the receiver,   which is challenging in practice \cite{ATP_fso}. Additionally, laser-based OWCs have several disadvantages, such as safety threats (e.g., hyperthermia and eye injury) and laser color mixing problems\cite{Chowdhury_optic}.
	
	Resonant beam communication (RBCom) is a new promising OWC approach to achieve high-speed data transmission while reducing the alignment difficulty between the transmitter and the receiver \cite{resonantcom, xiong_echo, xiong_harmonic}.    In the RBCom system,  two retroreflectors, which are located at the transmitter and the receiver, respectively, compose a  distributed resonator due to the self-alignment feature of the retroreflectors \cite{cornercube}. Then, the photons produced by the gain medium at the transmitter travel back and forth in this retroreflective resonator, forming a resonant beam connecting the transmitter and the receiver. Similar to the conventional OWC techniques, the resonant beam is modulated at the transmitter to carry information to the receiver. Moreover, the RBCom system has some advantages such as multiple access and self-alignment of the transceivers \cite{resonantcom, xiong_echo, xiong_harmonic}. Nevertheless, as the modulator works, the transmitted symbols running cyclicly in the resonator interfere the subsequent ones. This phenomenon is called echo interference \cite{resonantcom}.  The authors in \cite{xiong_echo} utilized frequency shifting and optical bandpass filtering to avoid the echo interference. However, due to the limitation of optical filters, the modulator has to operate the information signals with tens of GHz bandwidth, which is inefficient and difficult to be applied. In \cite{xiong_harmonic}, the authors proposed an alternative method for RBCom design, employing optical filtering and second harmonic generation (SHG) to eliminate the echo interference. However, the
	 practical non-ideal SHG medium significantly decreases the transmission efficiency, e.g., as indicated in \cite{ff_lu},  the SHG medium efficiency is typically only $2\%$. Additionally, complicated optical structures make it difficult to be implemented.

	The purpose of this two-part paper is to propose a more practical approach for the implementation of the RBCom system in both the quasi-static and mobile scenarios.
	In Part I of this paper, a quasi-static scenario is studied, where the transmitter and the receiver are relatively fixed.  Besides, the resonant beam moves back and forth between the transmitter and the receiver, and the transmitter starts to send information after it becomes stable.
	 Under this setup, 
	  the main contributions of Part I of  this paper are summarized as follows: 
	\begin{enumerate}
		\item  First, we propose a novel synchronization-based amplitude modulation method for the quasi-static RBCom system to mitigate the echo interference.  Specifically, the transmitter is designed to send one frame in each reflection round consisting of a synchronization sequence (SS) and  $N$ amplitude-modulated symbols, where symbol-level synchronization across different frames is achieved by the SS.
		\item  Then, based on the proposed method,   the RBCom channel is shown to be a Markov channel with infinite states, of which the link gain is analyzed by studying the properties of the gain medium and the propagation characteristics of the resonant beam. After that, an information-bearing scheme based on the reflected resonant beam is proposed, simplifying the Markov channel as an amplitude-constrained additive white Gaussian noise (AWGN) channel.
		\item Finally, based on the above analysis, we formulate the capacity upper and lower bounds maximization problems for the considered RBCom system. By utilizing the properties of the link gain, we propose an algorithm to approximately solve them.   Numerical results reveal that the capacity lower bound is in close proximity to the upper bound.
	\end{enumerate}
 The mobile scenario is more challenging and is studied in Part II of this paper \cite{Dong_part2}.
 
    The remainder of Part I of this paper is organized as follows. Section II describes the signal model and the synchronization-based amplitude modulation method. Section III simplifies the RBCom channel, and then solves the capacity upper and lower bounds maximization problems. Section IV presents the numerical results. Finally, Section V concludes this paper.   
    
    \emph{Notation}: $ \ln(\cdot) $ and $ \log_2(\cdot) $   denote the natural and  base-$2$ logarithms, respectively; $ \exp(\cdot) $ denotes the natural exponent;   $\lceil x \rceil $ denotes the smallest integer greater than or equal to $ x$; $\max\{x, y\}$ and $ \min\{x, y\} $ denote the maximum and  minimum between two real numbers $ x $ and $ y $, respectively.
 
   \section{System Model} \label{model}

\begin{figure}[htbp]
	\includegraphics[width=6in]{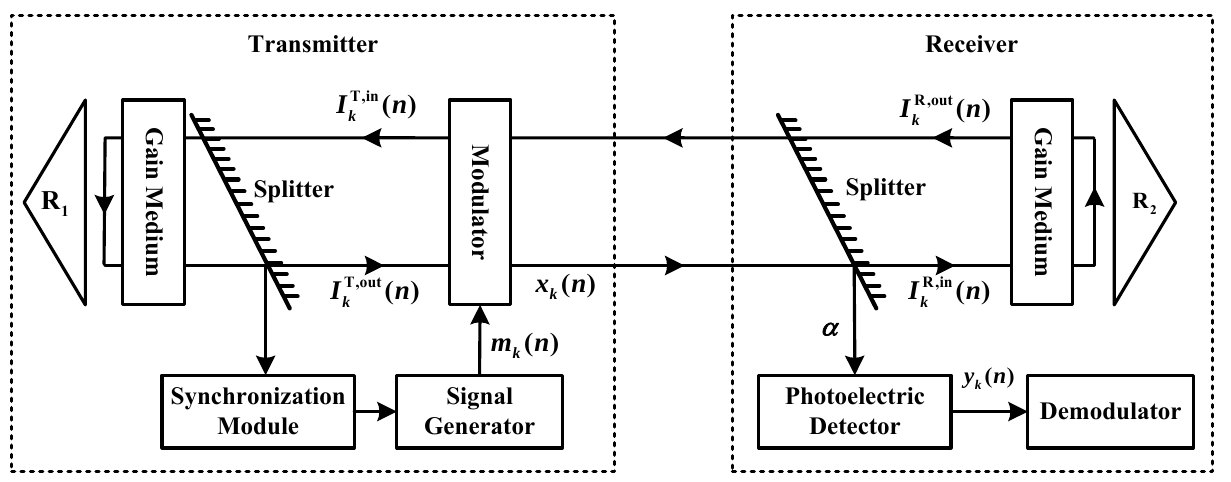}
	\caption{Structure of RBCom system.}
	\label{figure_system}
    \end{figure}

     As depicted in Fig. \ref{figure_system}, the considered RBCom system consists of one transmitter and one receiver, each of which is equipped with a retroreflector.  The resonant cavity is formed by retroreflectors $ \text{R}_1 $ and $ \text{R}_2 $, where the photons move cyclically to form a quasi-monochromatic resonant beam \cite{resonantcom}. 
      Before establishing communications, the resonant beam is amplified by two gain mediums until the gain equals the link loss in each reflection round. After the resonant beam becomes stable, i.e., the intensity approaches a constant over time, communication starts.
In each communication reflection round, the resonant beam at the transmitter is amplified by the gain medium, and then split by an optical splitter into two parts: One part is directed to the synchronization module to achieve the symbol-level synchronization by controlling the signal generator; the other part is sent to the free-space electro-optic amplitude modulator for carrying information. Without loss of generality, we consider the case that the power of the resonant beam split to the synchronization module is negligible compared to that of the modulator.  Besides, the modulator only works in one direction \cite{van2018model}, i.e., it only modulates the resonant beam passing from its left to right in Fig. \ref{figure_system} and has no effect on the beam traveling in the opposite direction.
At the receiver, the resonant beam is also split by an optical splitter into two beams: One beam with power ratio $ \alpha $ is transformed to the electrical signals by a photoelectric detector and then demodulated to recover the transmitted information; the other one with power ratio $ 1- \alpha$ is amplified by the gain medium at the receiver and reflected back to the transmitter by retroreflector $ \text{R}_2 $. 

  \subsection{Transmission at Transmitter} \label{1.a}
\begin{figure}[htbp]
	\centering
	\includegraphics[width=6in]{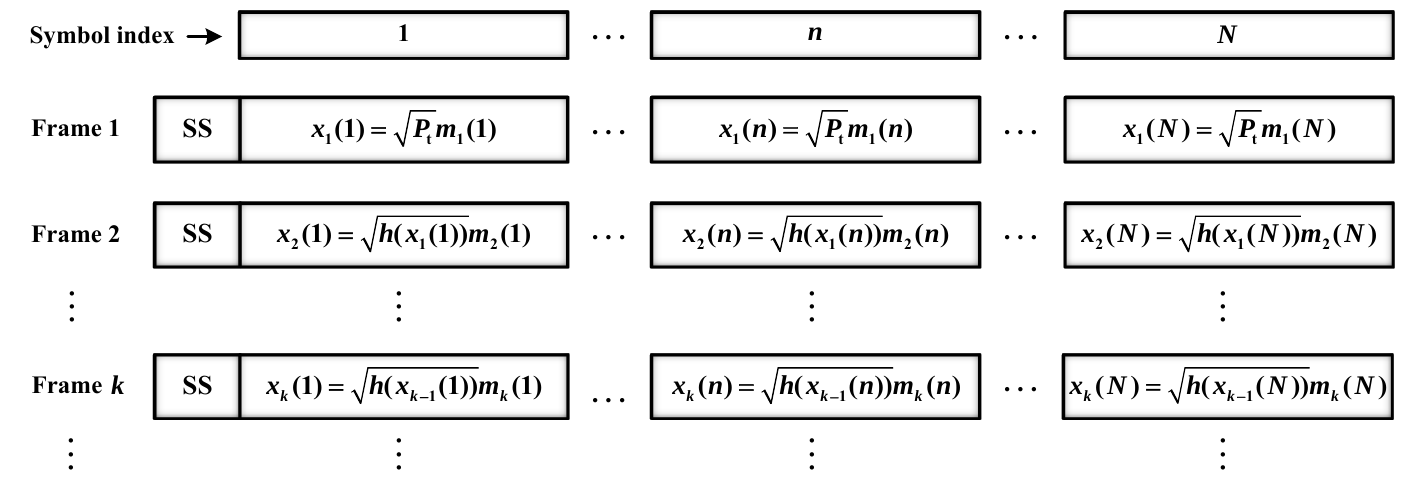}
	\caption{Data transmissions across different frames. }
	\label{synchro_symbol}
    \end{figure}

  We employ a frame-based transmission for the considered RBCom system where the length of each frame equals the duration of one reflection round of the resonant beam between the two retroreflectors. As shown in Fig. \ref{synchro_symbol}, the signals in each frame consist of an SS and $ N $ transmitted data symbols, where the SS ensures symbol-level synchronization across different frames \cite{ling2017synchronization}. Here, we denote $ m_k(n) $ as the  $ n$-th modulated symbol at frame $k$ which is produced by the signal generator at the transmitter, and $ x_k(n) $ as the $ n $-th transmitted symbol at frame $ k $.
   In the first reflection round, the transmitted symbol $ x_1(n) $ at frame 1 is produced by directly modulating $ m_1(n) $ onto the resonant beam and then transmitted to the receiver. In the $k$-th ($k\geq 2$) reflection round, frame $k-1$ is reflected back to the modulator, and the modulated symbol $ m_k(n) $ is modulated onto the reflected frame $ k-1 $ at the symbol level, which results in the generation of frame $k $ to be transmitted during the current round.  Therefore,  the $ n$-th transmitted symbol $x_k(n) $ at frame $ k $ is given as 
        \begin{equation}
    	x_k(n) =
    	\begin{cases}
    		\sqrt{P_\text{t}}m_k(n), \quad \quad & k = 1, \\
    		\sqrt{h(x_{k-1}(n))}m_k(n),& k = 2,3, \cdots,
    	\end{cases} 
    	 \quad  n = 1,2,\cdots, N,
    	\label{x_k}
    \end{equation}
	  where the modulated symbol $ m_k(n) $, $ 0 < m_k(n)\leq 1$\footnote{The modulator modulates the resonant beam by attenuating its intensity according to the modulated symbol $ m_k(n) $ \cite{damgaard2022electro}, where $ m_k(n) $ is not larger than 1. Moreover, $ m_k(n) $ cannot be 0 to avoid the interruption of the resonant beam.}, is an amplitude-constrained variable, and $ P_\text{t} $ is the stable power of the resonant beam before the starting of communications. Besides,  $ h(x_{k-1}(n)) $ is the link gain as a function of transmitted symbol $ x_{k-1}(n) $ in frame $ k-1 $, and it represents the instantaneous signal power after $ x_{k-1}(n) $ is reflected back to the transmitter, which jointly characterizes the amplification of the gain medium and the link loss between the transmitter and the receiver. The expression for function $ h(\cdot) $ will be discussed with more details later.
	\begin{Remark}
	  From \eqref{x_k}, it is observed that for any fixed $ n \in \{ 1,\cdots,N \} $,  transmitted symbol sequence $  \{x_k(n), k \geq 1 \} $ forms a discrete-time non-homogeneous Markov chain since the transition probabilities in \eqref{x_k} vary with $k$\cite{gallager2013stochastic}. Besides,  as shown in Fig. \ref{synchro_symbol}, any two symbol sequences $  \{x_k(n_1), k \geq 1 \} $ and $  \{x_k(n_2), k \geq 1 \} $, $ n_1\neq n_2 $, are independent when $ m_k(n)$'s are independent across $n$.
	  \label{Re_xk}
	\end{Remark}
	\subsection{Reception at Receiver}
\begin{figure}[htbp]
	\centering
	\includegraphics[width=4.7in]{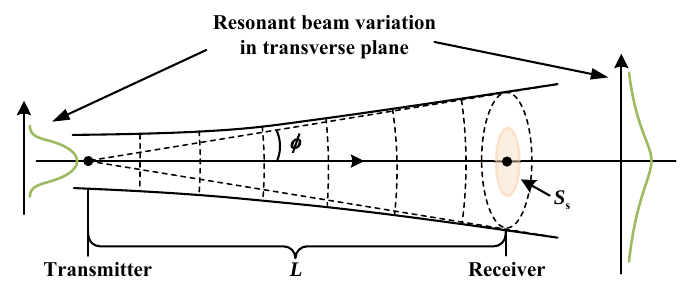}
	\caption{ Schematic diagram of resonant beam propagation from the transmitter to the receiver.}
	\label{fig_gaussain}
    \end{figure}

	Let $ \delta $ be the link loss \cite{lasercom} from the transmitter to the receiver. After being split by the splitter and detected at the photoelectric detector of the receiver,
	 the $n$-th  received symbol at frame $ k $  is given as 
     \begin{align}
      y_k(n)&= \sqrt{\alpha \delta}x_k(n) + \nu_k(n) \label{y_k_1} \\ 
       &=\begin{cases}
    		\sqrt{\alpha \delta P_\text{t}}m_k(n)+  \nu_k(n), & k = 1, \\
    		\sqrt{\alpha \delta h(x_{k-1}(n))}m_k(n)+ \nu_k(n),& k = 2,3,\cdots,
    	\end{cases} 
    	 \label{y_k}
     \end{align}       
     where  $ \nu_k(n) $'s are independent and identically distributed (i.i.d.) additive white Gaussian noise with mean zero and variance $ \sigma^2 $. For simplicity, we consider the case that the energy loss caused by the photoelectric detector is negligible. Fig. \ref{fig_gaussain} depicts the propagation of the resonant beam with a given diffraction angle $\phi$ from the transmitter to the receiver, with $L$ being the distance between the transmitter and receiver and $S_\text{s}$ being the receiving area of the receiver.  Since the spot area formed by the resonant beam at the receiver is larger than the receiving area $S_\text{s}$, the receiver can only receive part of the beam.   Considering the scenario that the resonant beam has a Gaussian intensity profile in the transverse plane \cite{longlaser}, link loss $ \delta $ from the transmitter to the receiver is obtained by the following proposition.

     \begin{Proposition} \label{prop_diffra}
   	   Considering the case that the intensity of the resonant beam follows Gaussian distribution \cite{laserbook1}, link loss $ \delta $   from the transmitter to the receiver   is given as 
   	   
   	   \begin{equation}
   	   	\delta = 1 - \exp\Bigg( \frac{-2S_\text{s}}{\frac{\lambda^2}{\pi\phi^2} + \pi\phi^2L^2} \Bigg),
   	   	\label{loss_define}
   	   \end{equation}
where  $ \lambda $ is the wavelength of the resonant beam.
    \end{Proposition}
    \begin{IEEEproof}
    Please see Appendix \ref{ap_loss_proof}.	
    \end{IEEEproof}
        
   \subsection{ Link Gain Function $ h(\cdot ) $}
   
\begin{figure}[t]
 	\centering
 	\includegraphics[width=4.5in]{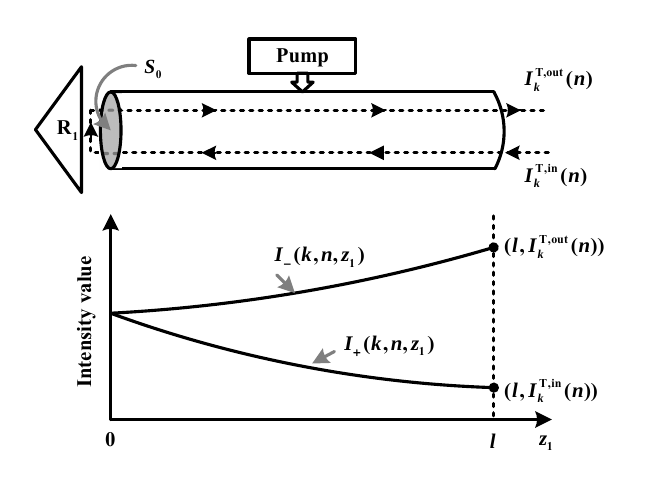}
 	\caption{Schematic diagram of the resonant beam in the gain medium at the transmitter \cite{Rigord1}. }   
 	\label{gain_part1}
    \end{figure}

   
    To derive the expression for link gain function $ h(x_k(n)) $,
 we first define power gain $G\big(I^{\text{T},\text{in}}_k(n)\big)  $ of the gain medium as the ratio of  output beam intensity  $ I^{\text{T}, \text{out}}_k(n) $ and input beam intensity $ I^{\text{T},\text{in}}_k(n) $ \cite{laserbook1}, i.e., 
    \begin{equation}
    	G\big(I^{\text{T},\text{in}}_k(n)\big) = \frac{I^{\text{T}, \text{out}}_k(n)}{I^{\text{T}, \text{in}}_k(n)}.
    	\label{G_defi}
    \end{equation}
    Fig. \ref{gain_part1} illustrates the variation of input intensity $ I^{\text{T},\text{in}}_k(n) $ along the $z_1$-axis as the resonant beam traverses through a rod-shaped\cite{resonantcom} gain medium at the transmitter. The gain medium has a length of $l$ and cross-sectional area $S_0=\pi r_0^2$, where $r_0$ is the radius of the circle obtained when the gain medium is sliced perpendicular to the $z_1$-axis.
    Furthermore, $ I_+(k,n,z_1)$ and $ I_-(k,n,z_1) $ are the incident and emergent beams of the gain medium, respectively. Assuming no reflection loss at the retroreflector $R_1$, it follows $ I_+(k,n,0) = I_-(k,n,0)$, $ I_+(k,n,l) = I_k^{\text{T},\text{in}}(n) $, and $ I_-(k,n,l) = I_k^{\text{T},\text{out}}(n) $. Then, utilizing Rigrod's formulation \cite{Rigord1}, we derive the following result.

    \begin{Proposition}
 \label{prop_I}
		   With given $ I^{\text{T}, \text{in}}_k(n) $, power gain $ G\big(I^{\text{T}, \text{in}}_k(n)\big) $ at the transmitter is uniquely obtained by the following equation
     \begin{equation}
     \label{G_calculate}
     	I^{\text{T}, \text{in}}_k(n) = \frac{2\eta P_{\text{in}}-I_\text{s}S_0 \cdot \ln
     	  G\big(I^{\text{T}, \text{in}}_k(n)\big)}{2 (G\big(I^{\text{T} \text{in}}_k(n)\big)-1)S_0},
     \end{equation}
     where $ I_\text{s} $, $ P_\text{in} $, and $ \eta $ are the saturation intensity \cite{laserbook1},  pumping power, and pumping efficiency of the gain medium at the transmitter, respectively.  
\end{Proposition}
    \begin{IEEEproof}
    Please see Appendix \ref{ap_prop_I}.	\end{IEEEproof} 

\begin{figure}[htbp]
	\centering
	\includegraphics[width=3.5in]{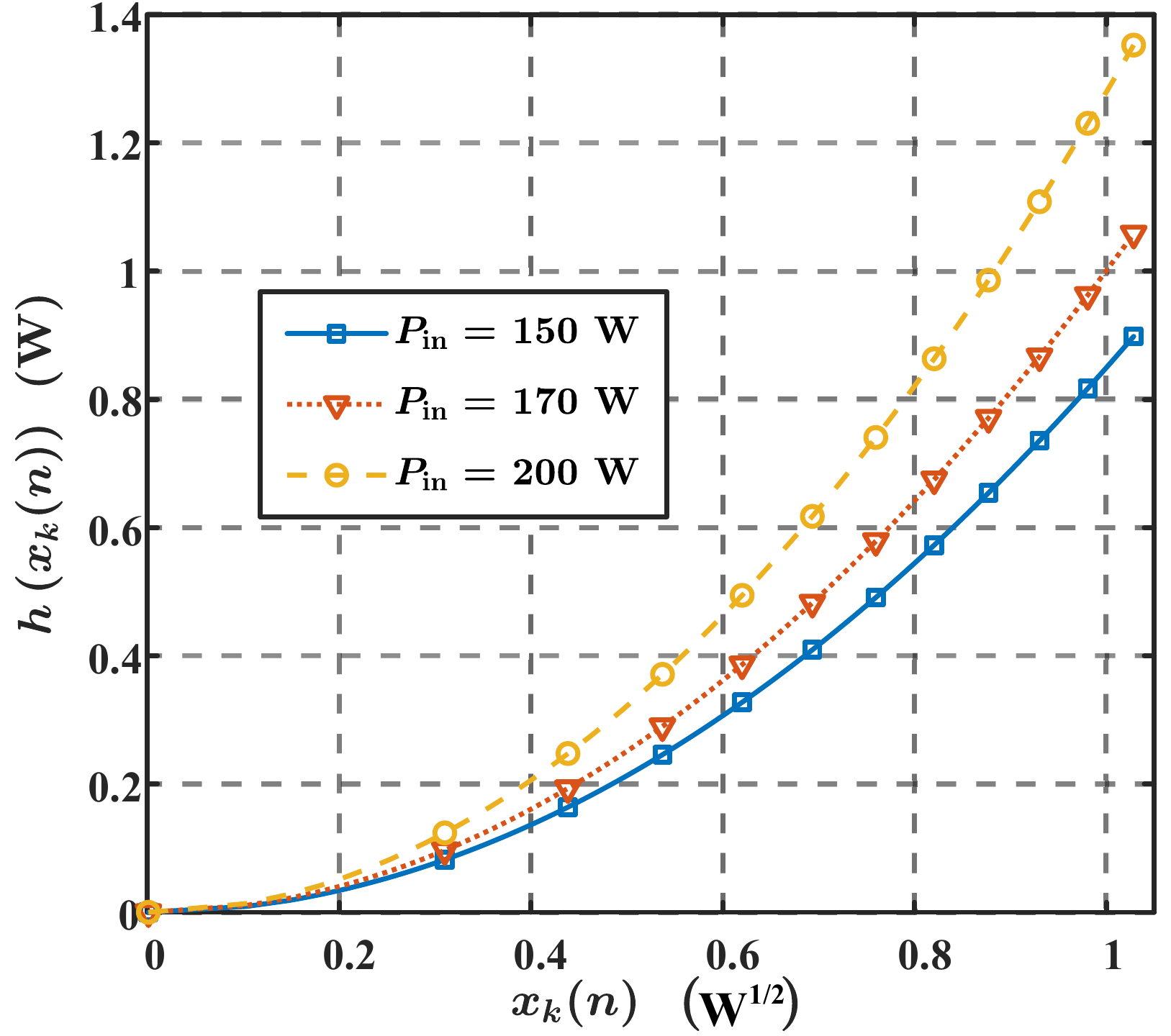}
	\caption{Link gain function $ h(x_k(n)) $ vs. transmitted symbol $ x_k(n) $. } 
	\label{h_figure}
    \end{figure}

       Then, we have the following proposition about link gain function $ h(\cdot) $.     
   \begin{Proposition}
   \label{prop_h}
   	When the gain mediums at the transmitter and the receiver are identical,   link gain function $ h(x_k(n)) $ in the $ k $-th frame is expressed as
   	 \begin{equation}
   	 \label{func_h}
   	 	h(x_k(n)) = (1-\alpha)\delta^2G\left( G\left(\frac{(1-\alpha)\delta x_k^2(n) }{S_0}\right)\frac{(1-\alpha)\delta^2 x_k^2(n) }{S_0}  \right) G\left(\frac{(1-\alpha)\delta x_k^2(n) }{S_0}\right)x_k^2(n).
   	 \end{equation}
   	 \end{Proposition}
   \begin{IEEEproof}
 	Please see Appendix \ref{ap_prop_h}.
   \end{IEEEproof}
\begin{Remark}  \label{re_h}
 	From the proof of Proposition \ref{prop_h}, we have shown that link gain function $ h(x_k(n)) $  monotonically increases as  $ x_k(n) $ increases, and satisfies  $h(x_k(n))< (1-\alpha)\delta^2e^{\frac{4\eta P_\text{in}}{I_\text{s}S_0}}x_k^2(n) $. Moreover, according to \eqref{func_h},  we draw the variation of $ h(x_k(n)) $ with respect to $ x_k(n) $ in Fig. \ref{h_figure} by selecting link loss $ \delta = 0.5 $, power splitting ratio $ \alpha = 0.01 $, setting parameters $ I_\text{s} $, $r_0$, and $ \eta $ of the gain medium as $ 1.2 \times 10^7 $ W/$ \text{m}^2 $ \cite{laserbook1}, $ 3 $ mm, and $ 0.7 $, respectively, and choosing $ P_\text{in} $ as $ 150 $, $ 170 $, and $ 200 $ W, respectively.  It is easy to observe that link gain function $h(x_{k}(n)) $ in \eqref{y_k} almost quadratically increases with respect to transmitted symbol $ x_k(n) $ and monotonically increases as $ P_\text{in} $ increases, which makes it difficult to analyze the capacity for the considered RBCom channel\cite{Aleksandar}.
 \end{Remark} 
 
\section{ Transmission  Rate Analysis} \label{Sec_AWGN}
In this section, we first propose an information-bearing scheme to simplify the channel of the considered RBCom as the amplitude-constrained AWGN channel. Then, we formulate the corresponding capacity upper and lower bounds maximization problems for the RBCom system, respectively. Finally, we develop an algorithm to compute these bounds.

\subsection{ Simplified Channel Model}  \label{sec_channel}
 To deal with the difficulties discussed in Remark \ref{re_h}, we design the modulated symbol $ m_k(n) $ in \eqref{y_k} to transform  $ \sqrt{h(x_{k-1}(n))} $ into a constant across all $ k $. Specifically, a straightforward approach to choose the modulated symbol $ m_k(n) $ is
\begin{equation}
	m_k(n) = w_k(n)s_k(n), \label{m_k}
\end{equation}
where $ s_k(n) $'s are i.i.d. information symbols for arbitrarily fixed $ n \in \{ 1,\cdots,N \}$, and $ w_k(n) \in(0,1] $ is designed to convert channel coefficient $ \sqrt{h(x_{k-1}(n))} $ to a constant, i.e.,
	\begin{equation}
	\label{w_k}
	\begin{cases}
	     \sqrt{P_\text{t}}w_k(n) = A_n, &k=1, \\   
		\sqrt{h(x_{k-1}(n))}w_k(n) = A_n, &k=2,3,\cdots,
	\end{cases}
	\end{equation}
 with $ A_n $ being the transformed constant channel coefficient for any given $ n\in \{ 1,\cdots,N \} $. 
However,  \eqref{w_k} may not always hold.  In the following, we establish the sufficient and necessary conditions for the validity of the design scheme \eqref{m_k}-\eqref{w_k}.
 \begin{Proposition}
\label{prop_wk}
	   The design scheme \eqref{m_k}-\eqref{w_k} holds if and only if $ s_k(n)$ is bounded over $[\mu_1(n), 1]  $ and its minimum value satisfies $ \mu_1(n) \geq \frac{\hat{A}_n}{A_n}  $, where $ \sqrt{h(\hat{A}_n)} = A_n $ and $ 0 <A_n \leq  \sqrt{P_\text{t}}$.
\end{Proposition}
 \begin{IEEEproof}
	 	Please see Appendix \ref{ap_prop_wk}.
\end{IEEEproof}

 \begin{Remark}
 	From Proposition \ref{prop_wk} and \eqref{m_k}, it is easy to see that $ s_k(n) $'s satisfying  $ s_k(n) \in [\mu_1(n), 1]$ are amplitude-constrained i.i.d. symbols. 
 	    Then, replacing $ m_k(n) $ in \eqref{x_k} with \eqref{m_k} and \eqref{w_k}, the transmitted symbol $ x_k(n) $ defined in \eqref{x_k} is simplified as
 	 \begin{equation}
 	 	x_k(n) = A_ns_k(n),\quad k =1, 2,3,\cdots,  \quad n = 1, \cdots, N. \label{simple_x_k}
 	 \end{equation}
 	  Here, it is obvious to observe that the transmitted symbol sequence $  \{x_k(n), k \geq 1 \} $ noted in Remark \ref{Re_xk} is no longer a non-homogeneous Markov chain. More specifically, for any fixed $ n $, the transmitted symbols $ x_k(n) $'s, $k\geq1$, are i.i.d., as $ A_n $ is a constant and $ s_k(n) $'s  are i.i.d. and amplitude-bounded. Moreover, replacing $ x_{k-1}(n) $ in \eqref{w_k} with \eqref{simple_x_k}, and together with \eqref{m_k},  
 	 modulated symbol $ m_k(n) $ is given as
    \begin{equation}
    \label{new_mk}
    	m_k(n) = \begin{cases}
    		\frac{A_ns_k(n)}{\sqrt{P_\text{t}}}, \quad    &k = 1, \\
    		\frac{A_ns_k(n)}{\sqrt{h(A_ns_{k-1}(n))}}, &k = 2,3, \cdots,
    	\end{cases} 
    \end{equation}
 	where  $ m_k(n) $ is jointly determined by information symbols $ s_k(n) $ at frame $ k $ and $ s_{k-1}(n) $ at frame $ k-1 $ for the case of $ k\geq 2 $.

 \end{Remark}
 \begin{Remark}
 	Beyond the design scheme  \eqref{m_k}-\eqref{w_k}, an alternative approach is to directly adjust pumping power $ P_\text{in} $ in real-time such that the variations of the link gain function  $ h(x_k(n)) $ in \eqref{y_k} correspond to the changes of the symbols being transmitted. However, this method is generally difficult to implement due to the strict requirements on the response time of the gain medium and pump source\cite{laserbook1}.
 \end{Remark}
 
  Together with  \eqref{simple_x_k}, the input and output model for RBCom  in \eqref{y_k} is rewritten as
     \begin{equation}
    y_k(n) = \sqrt{\alpha\delta}A_ns_k(n) + v_k(n),\quad k =1, 2,3,\cdots,  \quad n = 1, \cdots, N.
    	  \label{AWGN_model} 
    \end{equation}
    From \eqref{AWGN_model}, the channel of RBCom is now simplified as a group of  $ N $ parallel amplitude-constrained AWGN channels since $ A_n $ is a constant and $ s_k(n) $'s are amplitude-bounded and i.i.d. symbols for any fixed $ n $.
    
\begin{figure}[htbp]
	\centering
	\includegraphics[width=4in]{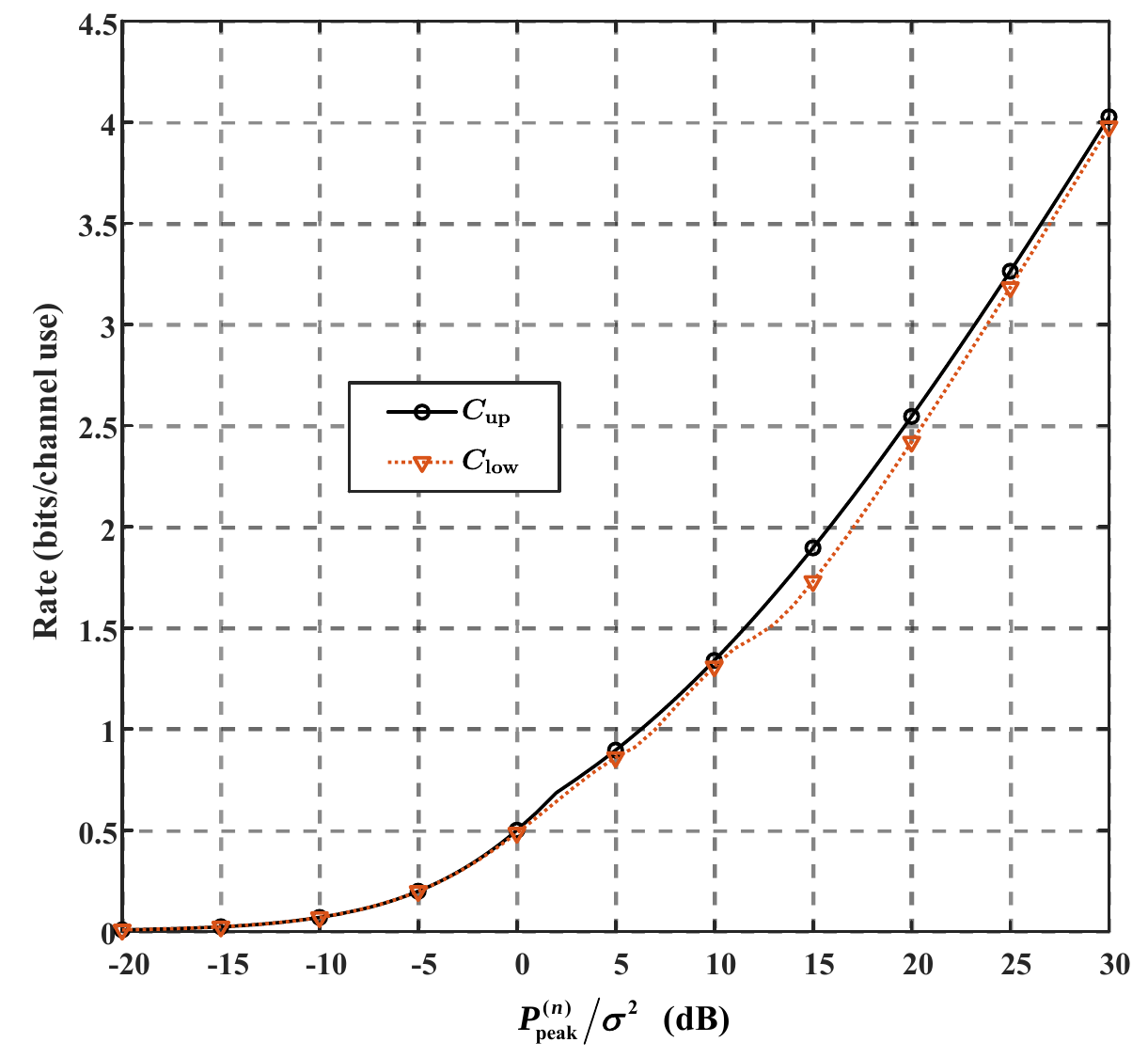}
	\caption{ Capacity upper and lower bounds for the amplitude-constrained AWGN  channel.} 
	\label{capacity_bound}
    \end{figure}

    Then, for the $n$-th amplitude-constrained AWGN channel, the   peak received signal power is thus given as \cite{smith} 
   \begin{equation}
   	  P_\text{peak}^{(n)} = \frac{(1-\mu_1(n))^2\alpha\delta A^2_n}{4}.
    	\label{P_peak}
   \end{equation}
   Capacity $C(P_\text{peak}^{(n)}) $ for the $n$-th amplitude-constrained AWGN channel was derived in \cite{smith}, while the computational resources required to compute $C(P_\text{peak}^{(n)}) $ are substantial \cite{Thangaraj}. Instead, we focus on the analysis of its capacity upper and lower bounds\cite{McKellips}, i.e.,   \begin{equation}
     C_{\text{up}}(P_\text{peak}^{(n)}) = \begin{cases}
		   \log_2 \left( 1 + \sqrt{\frac{2P_\text{peak}^{(n)}}{\pi e \sigma^2 }}  \right), &  \mathrm{if} \ P_\text{peak}^{(n)} >  \frac{ 8 \sigma^2 }{ \pi e ( 1 - \frac{2}{\pi e} )^2 },  \\
		\frac{1}{2}  \log_2\left(1+\frac{P_\text{peak}^{(n)}}{\sigma^2} \right),  &  \mathrm{otherwise},
	\end{cases}
     \label{cp_upper_bound}
    \end{equation}
and 
\begin{equation}
	 C_{\text{low}}(P_\text{peak}^{(n)}) = -\int_{-\infty}^{+\infty}f\left(y(n)\right)\log f\left(y(n)\right)dy(n)- \frac{1}{2}\log 2\pi e\sigma^2,
	 \label{cp_low_bound}
\end{equation}
respectively, where $ f(y(n)) $ is the probability density function of received symbol $ y(n) $ and generated by the distributions of information symbol $ s(n) $ and noise $ v(n) $ in \eqref{AWGN_model}, and $e$ is Euler's number. Meanwhile, $ s(n) $ follows a discrete uniform distribution on $ M $ equally-spaced points over interval $[\mu_1(n), 1] $ to achieve the capacity lower bound $  C_{\text{low}}(\cdot) $, and $ M $ is chosen as  \cite{McKellips}
\begin{equation}
	M(P_\text{peak}^{(n)}) = \begin{cases}
		2, & 0\leq\frac{P_\text{peak}^{(n)}}{\sigma^2} <2, \\
		3, & 2 \leq \frac{P_\text{peak}^{(n)}}{\sigma^2} <3.5, \\
		\left\lceil \frac{P_\text{peak}^{(n)}}{\sigma^2} \right\rceil, &\frac{P_\text{peak}^{(n)}}{\sigma^2} \geq 3.5.
	\end{cases} \label{C_low_M}
\end{equation}
 Fig. \ref{capacity_bound} plots capacity upper bound $ C_{\text{up}} $ and lower bound $ C_{\text{low}} $ for the amplitude-constrained AWGN channel of the RBCom system, which shows that  $ C_{\text{up}} $ and $ C_{\text{low}} $ are quite close for any given peak signal-to-noise ratio $ P_\text{peak}^{(n)}/\sigma^2 $. 
 
  As indicated by \eqref{P_peak}, the peak received signal power $  P_\text{peak}^{(n)}$ is affected by  $ \mu_1(n) $, $ A_n $ and splitting ratio $ \alpha $.
    In the following, by optimizing  $ \mu_1(n) $, $ A_n $, and $ \alpha $, we formulate and solve the maximization problems for $C_{\text{up}}(P_\text{peak}^{(n)}) $ and $C_{\text{low}}(P_\text{peak}^{(n)})$, respectively.

  \subsection{Upper Bound Optimization} 
  In this subsection, we first analyze the feasible region of $\alpha$.  Then, we formulate the maximization problem for $C_{\text{up}}(P_\text{peak}^{(n)})$ and jointly employ the bisection and exhaustive search to solve it.

	  \begin{Lemma}
  	  Given pumping power $ P_\text{in} $,	the feasible region of splitting ratio  $ \alpha $ to generate the resonant beam is given as
  	  	\begin{equation}
  	  	 \label{alpha_bound}
  	  		0 < \alpha < 1- \delta^{-2}e^{-\frac{4\eta P_\text{in}}{I_\text{s}S_0}}.
  	  	\end{equation} 
  	  \end{Lemma}
  	  \begin{IEEEproof}
  	  As stated in \eqref{x_k}, the resonant beam is stable with power $ P_\text{t} $ at the transmitter before the starting of communications, i.e., its power remains constant over time, which implies (by \eqref{G_calculate} and \eqref{func_h})
  	  \begin{equation}
  	  	h(\sqrt{P_\text{t}} ) =P_\text{t}. 
  	  	\label{stable_condition}
  	  \end{equation}
  	   Moreover, $ \lim_{P_\text{t} \to 0}\frac{h(\sqrt{P_\text{t}} )}{P_\text{t}} > 1  $ must be satisfied due to the monotonically decreasing property of function $ \frac{h(\sqrt{P_\text{t}} )}{P_\text{t}} $ given in  Lemma \ref{le_h} of Appendix \ref{ap_prop_h}. If not, $ h(\sqrt{P_\text{t}} ) < P_\text{t} $ always holds for any $P_\text{t} >0 $, which contradicts the stable condition \eqref{stable_condition} for the resonant beam. Then, together with \eqref{func_h} and Lemma \ref{le_I_convex} in Appendix \ref{ap_prop_I}, it follows $ \lim_{P_\text{t} \to 0}\frac{h(\sqrt{P_\text{t}} )}{P_\text{t}}= (1-\alpha)\delta^2\big(e^{\frac{2\eta P_\text{in}}{I_\text{s}S_0}}\big)^2 >1 $. Therefore, \eqref{alpha_bound} is  obtained.
  	  \end{IEEEproof}
  	  \begin{Remark}
  	  \label{Re_Pth}
  	  \eqref{alpha_bound} shows $  1- \delta^{-2}e^{-\frac{4\eta P_\text{in}}{I_\text{s}S_0}}  >0$, which is equivalent  to $ P_\text{in} > -\frac{I_\text{s}S_0\ln \delta }{2 \eta } $. Define $ P_\text{th} = -\frac{I_\text{s}S_0\ln \delta }{2 \eta } $ as the threshold pumping power for the RBCom system, which means that the resonant beam is produced if and only if $ P_\text{in} > P_\text{th} $. It has been shown that $ P_\text{th} $ is only determined by link loss $ \delta $ and parameters $ I_\text{s} $, $ S_0 $, and $\eta$ of the gain medium.
  	  \end{Remark}
    	  
    	  Then, the capacity upper bound maximization problem for the $n$-th amplitude-constrained  AWGN channel of the RBCom system is formulated as
  	   \begin{eqnarray}
   	\text{(P1)} & \max\limits_{ \{A_n, \mu_1(n), \alpha \} }   &C_{\text{up}}\left(\frac{(1-\mu_1(n))^2\alpha\delta A^2_n}{4}\right) \label{obj}  \\
  & \text{s.t.} &0\leq A^2_n\leq \min\left\{P_\text{t},\frac{P_\text{r,max}}{\alpha\delta}\right\}, \label{A_inequal}  \\
  &\quad & \frac{\hat{A}_n}{A_n} \leq \mu_1(n) \leq 1, \  \sqrt{h(\hat{A}_n)} = A_n, \label{mu_inequal}   \\
  &\quad & \eqref{alpha_bound}, \eqref{stable_condition},
   \end{eqnarray}
   where the objective function \eqref{obj} is derived by substituting $ P_\text{peak}^{(n)} $ with \eqref{P_peak} in \eqref{cp_upper_bound},   \eqref{A_inequal} is obtained from Proposition \ref{prop_wk} and  the maximum received signal power constraint, \eqref{mu_inequal}  is given in  Proposition \ref{prop_wk}, \eqref{alpha_bound} gives the feasible region of $ \alpha $, and \eqref{stable_condition} is utilized to compute $ P_\text{t} $ in  \eqref{A_inequal}.

    Furthermore, as shown in \eqref{P_peak}, $ P_{\text{peak}}^{(n)} $ monotonically increases  as  $ \mu_1(n) $ decreases. Then, the optimal solution to Problem ($ \text{P1} $) satisfies $  \mu_1(n) =\frac{\hat{A}_n}{A_n} $, since $ \frac{\hat{A}_n}{A_n} $  is the lower bound of $ \mu_1(n) $ and $  C_{\text{up}}(P_{\text{peak}}^{(n)})  $ monotonically increases with respect to  $ P_{\text{peak}}^{(n)} $. Therefore,  considering $  \mu_1(n) =\frac{\hat{A}_n}{A_n} $ and $  A_n=\sqrt{h(\hat{A}_n)}  $ by \eqref{mu_inequal}, Problem ($ \text{P1} $) can be equivalently simplified as   
  	   \begin{eqnarray}
   	\text{(P2)} \; &\max\limits_{ \{\hat{A}_n, \alpha\}} & C_{\text{up}}\left(\frac{(\sqrt{h(\hat{A}_n}) - \hat{A}_n)^2\alpha\delta }{4}\right) \label{obj_2}  \\
  	 & \text{s.t.} 
& 0\leq h(\hat{A}_n)\leq \min\left\{P_\text{t},\frac{P_\text{r,max}}{\alpha\delta}\right\}, \label{hat_A_inequal}  \\
&  & \eqref{alpha_bound}, \eqref{stable_condition}, \label{P_simple_ine} 
   \end{eqnarray}   
  where \eqref{obj_2} is obtained from \eqref{obj} by substituting $ \mu_1(n) $ with $ \frac{\hat{A}_n}{A_n} $, and \eqref{hat_A_inequal} is derived from  \eqref{A_inequal} by replacing $A_n$ with $\sqrt{h(\hat{A}_n)}$. 
  
  However, there are several difficulties in solving Problem (P2): First, from \eqref{func_h} and \eqref{stable_condition}, it is noted that  $ P_\text{t} $ in \eqref{hat_A_inequal} is an implicit function of $ \alpha $. Besides, for a fixed $ \alpha $, it is difficult to compute $ P_\text{t} $ directly since the expression \eqref{func_h} for $ h(\cdot)$ is quite complicated; Second,  Problem (P2) is generally non-convex due to the non-convex nature of both the objective function \eqref{obj_2} and the constraint \eqref{hat_A_inequal}. To deal with these difficulties, we first adopt the bisection search to effectively compute $ P_\text{t} $. Specifically, we derive the upper and lower bounds for $ {P_\text{t}} $ in the following lemma.  
   \begin{Lemma}
   \label{le_Pt}
   	Stable transmitted power $ P_\text{t} $ is bounded by
   	\begin{equation}
   	\label{P_cal}
   		 \max \left\{0, \frac{2\eta P_\text{in} + ( \ln \delta + \ln (1-\alpha) )I_\text{s}S_0  }{2(1-(1-\alpha)\delta )}\right\} \leq P_\text{t} \leq \frac{2\eta P_\text{in} + I_\text{s}S_0\ln \delta  }{2(1-\delta )} .
   	\end{equation}
   	
   	\begin{IEEEproof}
   	Please see Appendix \ref{ap_le_Pt}.
    	\end{IEEEproof}
   \end{Lemma}
   
   By exploiting the monotonically decreasing and continuous properties of function $  \frac{h(\sqrt{P_\text{t}} )}{P_\text{t}} $ given in Lemma \ref{le_h} of Appendix \ref{ap_prop_h}, and in combination with \eqref{stable_condition} and \eqref{P_cal},  we can utilize the bisection search to approximately compute $P_\text{t}$ for any fixed $ \alpha $.  
   
   Then, given that there are only two bounded optimization variables $ \hat{A}_n $ and $ \alpha $ in Problem (P2), we utilize the exhaustive search to efficiently solve this problem by discretizing $ \hat{A}_n $ and $ \alpha $.

  \begin{Remark}
  \label{Re_total_rate} From Problem (P2), it is easy to observe that the objective function in \eqref{obj_2} and constraints \eqref{hat_A_inequal}-\eqref{P_simple_ine} do not change with $ n $.  Therefore, 
  	    the optimal solution to  Problem (P2) is independent of $n$, which means that $ A_n $'s in \eqref{AWGN_model} can be designed as the same value to maximize the capacity upper bound of the considered RBCom system. 
  \end{Remark}

  \subsection{Lower Bound Optimization}
   
Substituting  $C_{\text{up}}(\cdot) $ with $ C_{\text{low}}(\cdot)$ in Problem (P2), the capacity lower bound maximization problem for the $n$-th amplitude-constrained  AWGN channel of the RBCom system is formulated as
  \begin{eqnarray}
     	\text{(P3)}  &\max\limits_{ \{\hat{A}_n, \alpha\}} & C_{\text{low}}\left(\frac{(\sqrt{h(\hat{A}_n}) - \hat{A}_n)^2\alpha\delta }{4}\right) \label{P_C_low}  \\
  &\text{s.t.} 
  &\eqref{alpha_bound},\eqref{stable_condition},  \eqref{hat_A_inequal}.	
  \end{eqnarray}
  Apparently, the optimal solution for Problem (P2)  is also optimal for Problem (P3), since $ C_{\text{up}} $ defined in \eqref{cp_upper_bound} is monotonically increasing with respect to its input $ \frac{(\sqrt{h(\hat{A}_n}) - \hat{A}_n)^2\alpha\delta }{4} $ and $ C_{\text{low}} $ defined in \eqref{cp_low_bound} is monotonically nondecreasing with respect to its input $ \frac{(\sqrt{h(\hat{A}_n}) - \hat{A}_n)^2\alpha\delta }{4} $. Moreover, similar to the upper bound optimization, it can be easily deduced that the optimal solution to Problem (P4) is also independent of $n$.

 \subsection{Algorithm}   
Based on the above analysis, we propose an algorithm, summarized as Algorithm \ref{alg_rate}, for computing the capacity upper and lower bounds of the considered RBCom system. In this algorithm, we first discretize the feasible regions of the bounded variables $ \hat{A}_n $ and $ \alpha $ into the discrete sets, respectively, and then jointly employ the bisection search and exhaustive search to obtain the optimal values $C_\text{up}^*$, $C_\text{low}^*$, $\alpha^*$, $\hat{A}^*$, and $ P_\text{t}^* $. Here, $\hat{A}^*$ is denoted as the optimal value of $ \hat{A}_n $ for all $ n =1,\cdots, N $ since we have shown that Problem (P2) and (P3) are both independent of $n$.  Besides, the optimal design of the modulated symbol $m_k(n)$ in \eqref{x_k} is also derived in Algorithm \ref{alg_rate}, accordingly.

    \begin{algorithm}[htbp]

	\caption{Bisection and exhaustive search for the capacity upper  and lower bounds of the RBCom system.\label{alg_rate} }
	\label{Joint Parameter Estimation And Data Detection} 
	\begin{algorithmic}[1]
	\Require  $ \lambda $, $ I_\text{s} $, $ \eta $, $S_\text{s}$,  $ S_0 $, $\phi$, $ \sigma^2 $, $B$, $ P_\text{in}$, and $ P_\text{r, max} $.
	\Ensure $C_\text{up}^*$, $C_\text{low}^*$, $\alpha^*$, $\hat{A}^*$, $ P_\text{t}^* $, and $m_k^*(n)$. 
    \State Compute $ \delta $ by \eqref{loss_define};
    \State Set $[C_\text{up}^* , C_\text{low}^* ,\alpha^*, \hat{A}^*, P_\text{t}^* ] = [0, 0,0, 0, 0] $ and $ K_1 =K_2 = 1000 $;
	    \State Compute the threshold power $ P_\text{th}  = -\frac{I_\text{s}S_0\ln \delta }{2 \eta }  $ based on Remark \ref{Re_Pth};
    \State Set pumping power $ P_\text{in} $  satisfying  $ P_\text{in} >  P_\text{th}  $; 
    \State Obtain the discrete set of $ \alpha :\mathbb{S}_{\alpha}= \{ \frac{1}{K_1},\frac{2}{K_1},\cdots,\frac{K_1-1}{K_1} \}\cdot( 1- \delta^{-2}e^{-\frac{4\eta P_\text{in}}{I_\text{s}S_0}})$ by \eqref{alpha_bound};
    \State \textbf{For} $ k_1= 1,2,\cdots, K_1-1 $ \textbf{do}
    \State \quad \ \ Set $ \alpha $  as $ \alpha = \frac{k_1}{K_1}\cdot( 1- \delta^{-2}e^{-\frac{4\eta P_\text{in}}{I_\text{s}S_0}}) $;
    \State \quad \ \ Utilize bisection search to compute $ P_\text{t}$ by \eqref{stable_condition} and \eqref{P_cal};
    \State \quad \ \ Obtain the discrete set of $ \hat{A}_n: \mathbb{S}_{\hat{A}_n}=  \{ \frac{1}{K_2},\frac{2}{K_2},\cdots,\frac{K_2-1}{K_2} \} \cdot \min\{P_\text{t},\frac{P_\text{r,max}}{\alpha\delta}\} $ by \eqref{hat_A_inequal};
    \State \quad \ \ \textbf{For} $ k_2= 1,2,\cdots, K_2-1 $ \textbf{do}
    \State \quad \quad \ \ \ \ Set $ \hat{A}_n $  as  $ \hat{A}_n = \frac{k_2}{K_2} \cdot \min\{P_\text{t},\frac{P_\text{r,max}}{\alpha\delta}\} $;
     \State \quad \quad \ \ \ \  Compute $ C_\text{up} $ and $C_\text{low}$ by \eqref{obj_2} and \eqref{P_C_low}, respectively;
     \State \quad \quad \ \ \ \ \textbf{If} $ C_\text{up} > C_\text{up}^* $ \textbf{then}
     \State \quad \quad \ \ \ \ \quad \ \ $[C_\text{up}^* , C_\text{low}^* ,\alpha^*, \hat{A}^*, P_\text{t}^* ] \leftarrow [C_\text{up}, C_\text{low},\alpha, \hat{A}_n, P_\text{t}] $;
      \State \quad \quad \ \ \ \ \textbf{End if} 
     \State \quad \ \ \textbf{End for}
    \State \textbf{End for}
    \State Compute $ A^* =\sqrt{h(\hat{A}^*)}  $ with $ \alpha = \alpha^* $ and produce the i.i.d. information symbol $ s_k^*(n)\in[\frac{\hat{A}^* }{A^*},   1] $ satisfying its optimal distribution given in \cite{smith};
     \State Compute the optimal modulated symbol $ m_k^*(n) $ by \eqref{new_mk}, where $ A_n $, $ P_{\text{t}} $, and $ s_k(n) $ in  \eqref{new_mk} are replaced with $ A^* $, $P_\text{t}^*$, and $ s_k^*(n) $, respectively.
	\end{algorithmic}
\end{algorithm}

 \section{Numerical Results}
 
 In this section, we present some numerical results to validate our theoretical results.  For the purpose of exposition, an Nd:YAG rod\cite{laserbook1} is adopted as the gain medium. Besides, radius $ r_0$ of the gain medium is set as  $3$ or $5$ mm, and accordingly,  $ S_0 =\pi r_0^2=28.26$ or $78.5$ $\text{mm}^2 $. For simplicity, the cross-sectional area of the gain medium has been set to be equal to that of the receiving surface at the receiver, i.e., $ S_0 = S_\text{s} $. Moreover, parameters  $ I_\text{s} $ and $ \eta $ of the gain medium are set the same as those in Remark \ref{re_h}.  The maximum power $ P_\text{r,max} $  of the received signal at the photodetector is set as $ P_\text{r,max} = 10 \ \text{dBm} $ and the wavelength of the resonant beam is set as $ \lambda = 1064$ nm. Besides, the bandwidth of the channel is fixed at $ B $ = $1$ GHz, and the power spectral density of the noise is set as $ N_0 =  -174 \ \text{dBm/Hz}  $. Therefore, noise power $ \sigma^2 = N_0B = -84$ dBm.  Moreover, two diffraction angles, $\phi = 0.2$ mrad and $\phi = 0.3$ mrad, have been considered for the subsequent analysis.

\begin{figure}[htbp]
	\centering
	\includegraphics[width=3.5in]{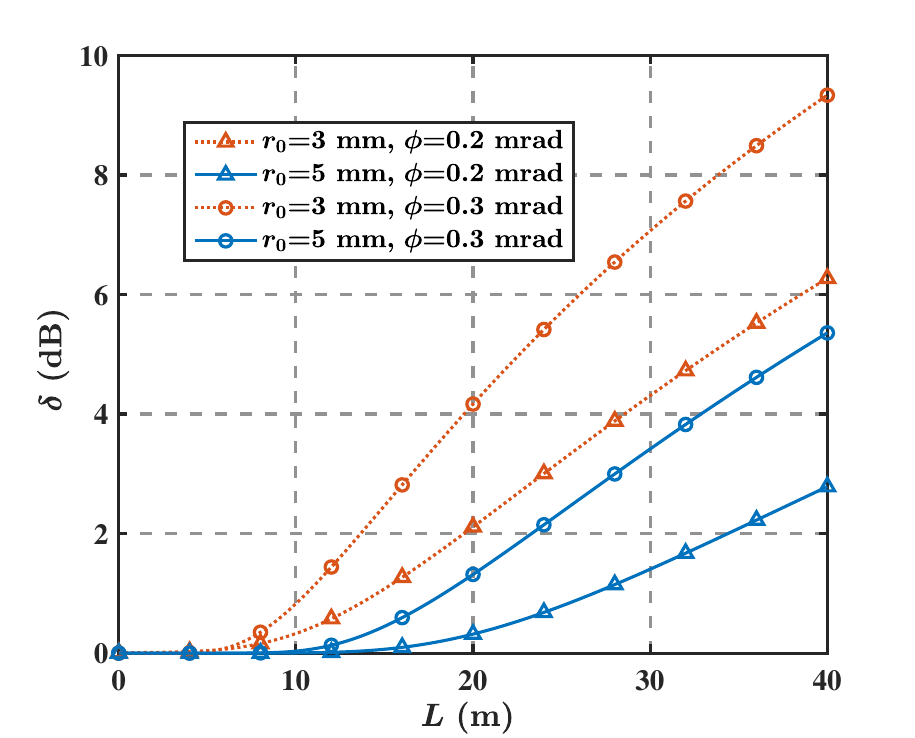}
	\caption{ Link loss  $ \delta $ vs. distance $ L $.  }
	\label{fig_L_delta}
    \end{figure}

  Fig. \ref{fig_L_delta} describes link loss $ \delta $  as a function of distance $ L $ between the transmitter and the receiver for different $ r_0 $ and $ \phi $. It is observed that $ \delta $  increases slowly for small values of $L$, e.g., $ L <10 $ m, and increases asymptotically exponentially as $L$ becomes larger, e.g.,  $ L > 15 $ m. Besides, $ \delta $ is dramatically affected by parameters $ r_0 $ and $ \phi $ when $ L $  is larger than $ 10 $ m. It is also noted that, for any fixed $L > 10$ m, $ \delta $ increases as $ r_0 $ decreases. Moreover, the increase of $\phi$ results in the corresponding increase of the value $\delta$.

\begin{figure}[htbp]
	\centering
	\includegraphics[width=3.5in]{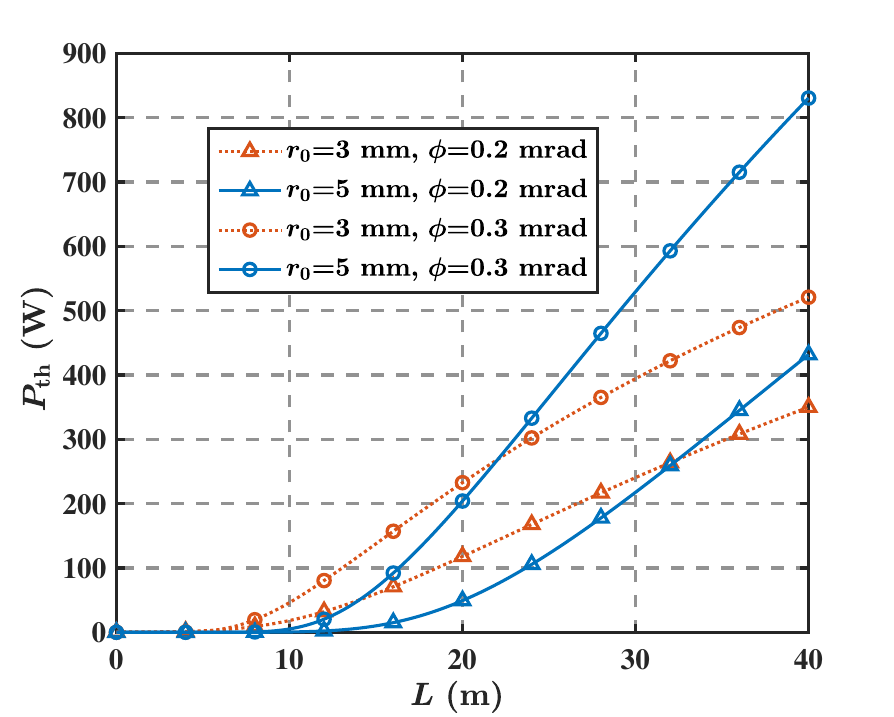}
	\caption{Threshold pumping power $ P_\text{th} $ vs. distance $ L $. }
	\label{fig_L_Pth}
    \end{figure}

    In Fig. \ref{fig_L_Pth}, based on Remark \ref{Re_Pth},  we draw threshold pumping power $ P_\text{th} $ as a function of distance $ L $ for different $ r_0 $ and $ \phi $.  It shows that $ P_\text{th} $ monotonically increases as $ L $ increases for any fixed $ r_0 $ and $ \phi $. Then, for any fixed $ L $ and $ r_0 $,  the value of $ P_\text{th} $ remains consistently lower in the scenario with $\phi = 0.2$ mrad compared to that with $\phi = 0.3$ mrad.  Moreover, with a constant value of $ \phi $, the increase in $L$ results in a notably higher growth rate of $P_\text{th}$ in the scenario with $r_0=5$ mm when compared to the case with $r_0=3$ mm.
   
  \begin{figure}[htbp]
	\centering
	\includegraphics[width=3.5in]{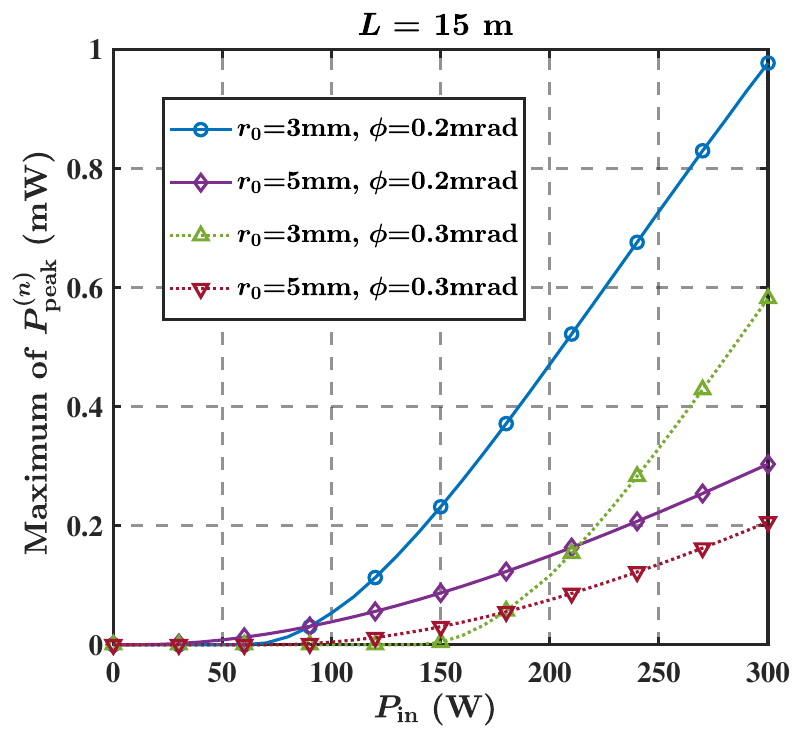}
	\caption{Maximum of the peak received signal power $ P_\text{peak}^{(n)} $ vs.  pumping power $ P_\text{in} $. }
	\label{fig_peak_power} 
    \end{figure}  
In Fig. \ref{fig_peak_power}, we draw the maximum value of $ P_\text{peak}^{(n)} $ as a function of the pumping power $ P_\text{in} $ for $L = 15$ m, utilizing the expression in \eqref{P_peak} and the results obtained from Algorithm \ref{alg_rate}.   Compared to Fig. \ref{fig_L_Pth}, it becomes apparent that $P_\text{peak}^{(n)}$ equals zero if $P_\text{in} \leq P_\text{th}$, as this condition implies that the resonant beam can not be produced. Furthermore, $P_\text{peak}^{(n)}$ exhibits a monotonic increase for $P_\text{in} > P_\text{th}$. Besides, it is noteworthy that a larger diffraction angle $\phi$ results in a smaller maximum value of $P_\text{peak}^{(n)}$. Moreover, with a constant value of $ \phi $, the increase in $P_\text{in}$ results in a  higher growth rate of the maximum of $P_\text{peak}^{(n)}$ in the scenario with $r_0=3$ mm when compared to the case with $r_0=5$ mm.
     
     \begin{figure}[htbp]
	\centering
	\includegraphics[width=3.5in]{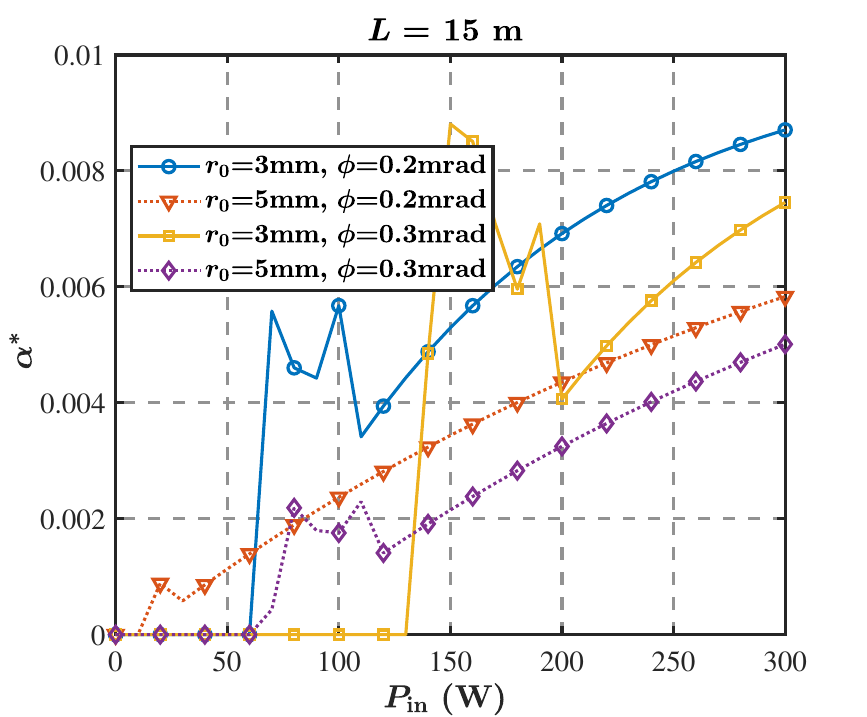}
	\caption{ Optimal splitting ratio $ \alpha^* $ vs.  pumping power $ P_\text{in} $. }
	\label{fig_Pin_alpha}
    \end{figure} 
    
     \begin{figure}[htbp]
	\centering
	\includegraphics[width=3.5in]{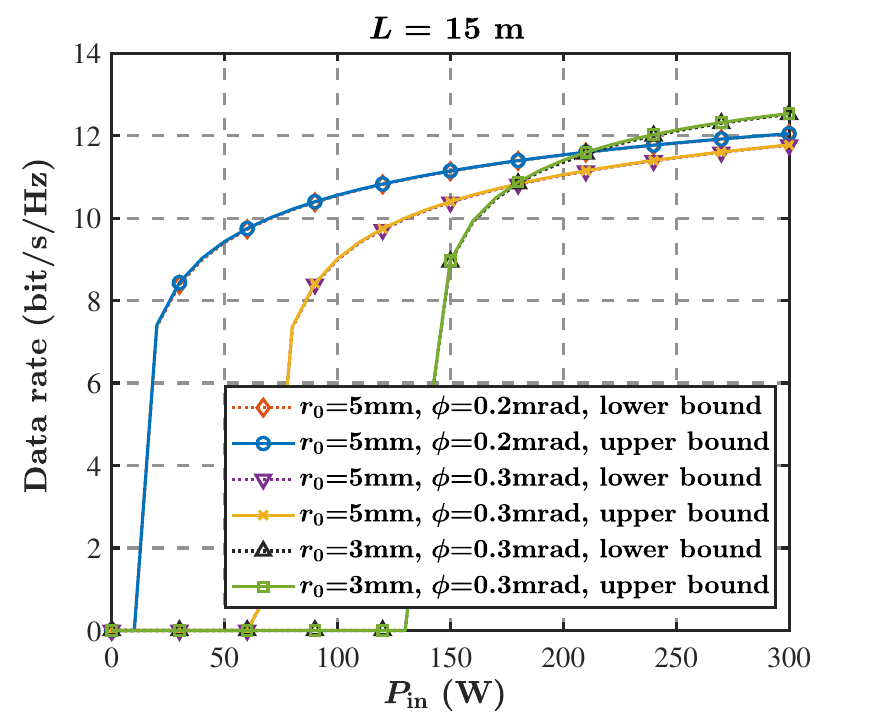}
	\caption{ Comparison of the capacity  upper and lower bounds for different $ P_\text{in} $, $ r_0 $,  and  $ \phi $.}
	\label{fig_Pin_rate}
    \end{figure}

 Fig. \ref{fig_Pin_alpha} describes the relationship between the optimal splitting ratio $ \alpha^* $  and pumping power $ P_\text{in} $ for different parameters $ r_0 $ and $ \phi $.   Similar to Fig. \ref{fig_peak_power}, when $P_\text{in} \leq P_\text{th}$, $\alpha^* = 0$, and for $P_\text{in} > P_\text{th}$, $\alpha^*$ remains small ($\alpha^* < 0.01$). Besides, when $P_\text{in}$ is sufficiently large, $ \alpha^* $ monotonically increases with respect to $P_\text{in}$.      
       \begin{figure}[htbp]
	\centering
	\includegraphics[width=3.5in]{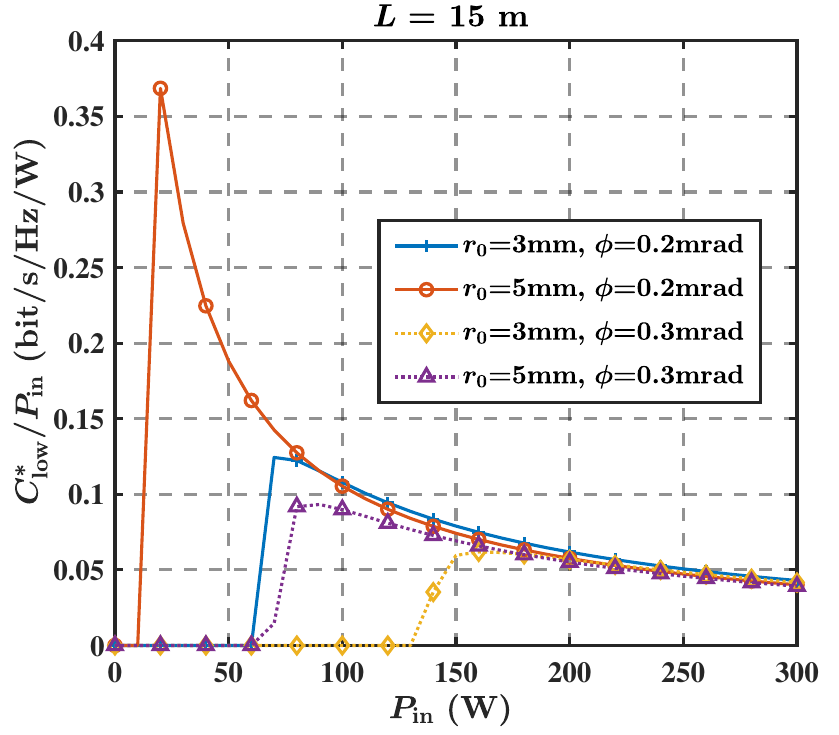}
	\caption{ $ C_\text{low}^*/P_\text{in}$ vs. pumping power $ P_\text{in}$.}
	\label{fig_Pin_effi}
    \end{figure}
    
    \begin{figure}[htbp]
	\centering
	\includegraphics[width=3.5in]{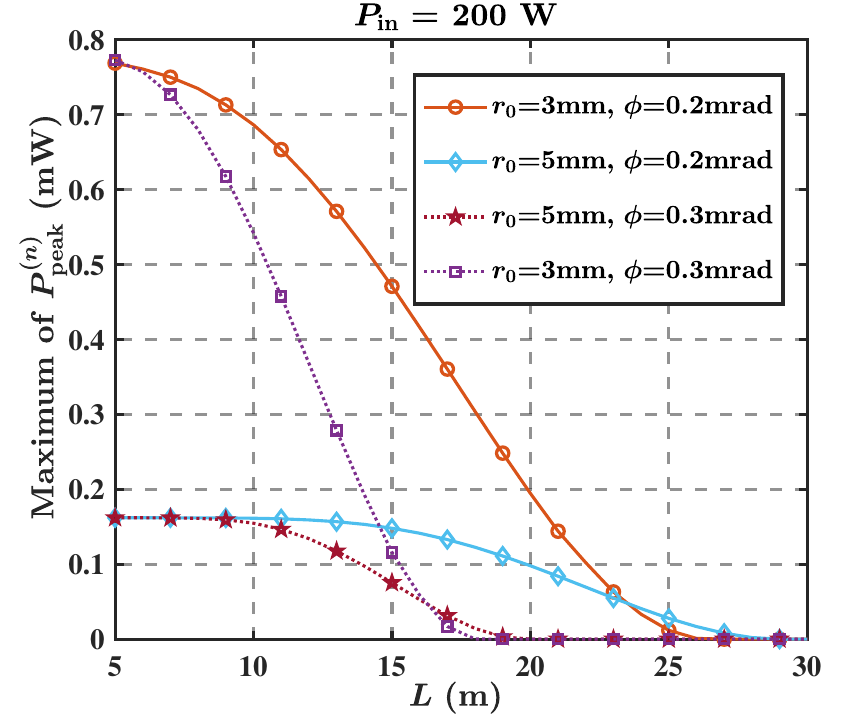}
	\caption{ 
	Maximum of $ P_\text{peak}^{(n)} $ vs. distance $ L $.  }
	\label{fig_L_Ppeak}
    \end{figure}

        	\begin{figure}[htbp]
	\centering
	\includegraphics[width=3.5in]{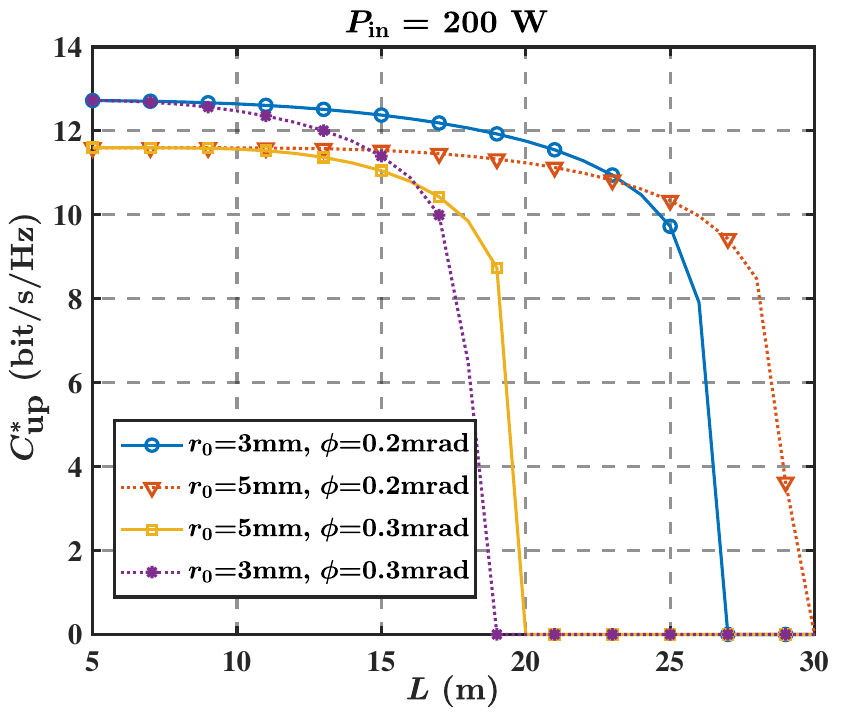}
	\caption{  Capacity upper bound $ C_\text{up}^* $  vs. distance $ L $.}
	\label{fig_L_rate}
    \end{figure}

    Based on Algorithm \ref{alg_rate}, Fig. \ref{fig_Pin_rate} describes the capacity upper and lower bounds $ C_\text{up}^* $ and $ C_\text{low}^* $  as functions of pumping power $ P_\text{in} $ for different parameters $ r_0 $ and $ \phi $. The upper and lower bounds of capacity are found to be nearly indistinguishable. Besides, the data rate  is $0$ if $ P_\text{in} \leq P_\text{th} $, and it increases dramatically if $ P_\text{in} > P_\text{th} $.
     Then, Fig. \ref{fig_Pin_effi} shows $ C_\text{low}^*/P_\text{in}$ as a function of pumping power $ P_\text{in} $, for four cases of $r_0$ and $\phi$, which shows the variations of the lower bound of energy efficiency. It is observed that  $ C_\text{low}^*/P_\text{in}$ initially increases and then steadily decreases as $P_\text{in}$ increases.  

In Fig. \ref{fig_L_Ppeak}, we plot the maximum of $ P_\text{peak}^{(n)} $ as functions of distance $ L  $ for fixed pumping power $ P_\text{in} = 200 $ W and different parameters $ r_0 $ and $ \phi $. It shows that the maximum of $P_\text{peak}^{(n)} $ monotonically decreases as $ L $ increases. Notably, the maximum value of $P_\text{peak}^{(n)}$ in the scenario where $r_0 = 3$ mm significantly surpasses that in the case with $r_0 = 5$ mm when $ L $ is quite small, e.g., $ L < 17 $ m and $ \phi=0.3 $ mrad, and the former decreases much faster to $0$ as $ L $ increases. Similarly, for a fixed $ r_0 $,  it is also observed that the maximum of $P_\text{peak}^{(n)}  $ in the case with $ \phi = 0.2  $ mrad is nearly identical to that with $ \phi = 0.3  $ mrad when $ L $ is close to $5$ m. This occurs because, as shown in Fig. \ref{fig_L_delta}, link loss $\delta$ approaches $0$ dB if $\phi=0.2$ or $0.3$ mrad and $L \leq 5$ m.  Based on these observations, we present in Fig. \ref{fig_L_rate} the capacity upper bound $ C_\text{up}^* $ as a function of distance $ L $ for fixed pumping power $ P_\text{in} = 200 $ W, which shows that the capacity upper bound remains relatively constant for $ L $ values below $15$ m but drops sharply as $ L $ increases.

    \section{Conclusion}
    
    In this paper,  we studied a point-to-point RBCom system under the quasi-static scenario with the transmitter and the receiver being relatively stationary. Specifically,  we proposed a synchronization-based amplitude modulation method to eliminate the echo interference and simplify the channel of RBCom as an amplitude-constrained AWGN channel.  After that, we formulated the non-convex capacity upper and lower bounds maximization problems for the RBCom system and leveraged bisection and exhaustive search to solve them. Numerical results demonstrated the tightness of the bounds, with the capacity upper bound being remarkably close to its lower bound. The mobile case is treated in Part II \cite{Dong_part2}.

  \bibliographystyle{IEEEtran}   
  \bibliography{model_resonant} 
\appendices

  \section{Proof of Proposition \ref{prop_diffra}} \label{ap_loss_proof}
	The diffraction of the resonant beam emitted from the transmitter results in the formation of a light spot at the receiver\cite{laserbook1}. Then, 
	in the $ k $-th reflection round, the intensity distribution $ I_{k,n}(r) $ on this light spot is defined as\cite{Siegman} 
	\begin{equation}
		I_{k,n}(r) = \frac{2x_k^2(n)}{\pi \omega^2} \exp \left[-\frac{2r^2}{\omega^2}\right],
		\label{ap_intensity}
	\end{equation}
	and 
	\begin{equation}
		\omega^2 = \omega_0^2\left[1+\left(\frac{\lambda L}{\pi\omega_0^2}\right)^2\right],
		\label{ap_omega}
	\end{equation}
	where $ r $ is the distance to the center point of the light spot, and $ \omega_0 $ is the waist radius of the resonant beam \cite{laserbook1}. The area of the circular receiver retroreflector is given as $S_\text{s}=\pi r_\text{s}^2$, with $r_\text{s}$ denoting the radius of the retroreflector. Then, link loss $\delta $ in free space is given as 
	\begin{equation}
		\delta = \frac{\int_0^{2\pi}\int_0^{r_\text{s}}I_{k,n}(r)rdrd\theta }{x_k^2(n)}=1- \exp \left( \frac{-2S_\text{s}}{\pi\omega^2} \right) .
		\label{ap_loss}
	\end{equation}	 
 Moreover, diffraction angle $\phi$ is given as \cite{laserbook1}
\begin{equation}
\label{ap_angle}
	\phi = \frac{\lambda}{\pi \omega_0}.
\end{equation}
Then, \eqref{loss_define} is obtained by incorporating \eqref{ap_intensity}, \eqref{ap_omega}, \eqref{ap_loss}, and \eqref{ap_angle}.

\section{Proof of Proposition \ref{prop_I} }
 \label{ap_prop_I} 
 
   \begin{figure}[htbp]
	\centering
	\includegraphics[width=3in]{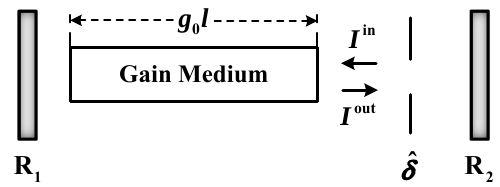}
	\caption{  Schematic diagram of a classic laser cavity with ideal retroreflectors ($ \text{R}_1 $ and $ \text{R}_2 $) and total energy loss $ \hat{\delta} $.} 
	\label{laser_system}
    \end{figure} 
To prove this proposition,  we present a schematic diagram of the classic laser cavity shown in Fig. \ref{laser_system},  with the gain medium parameters identical to those depicted in Fig. \ref{gain_part1}. For simplicity, we denote the input and output stable intensities of the gain medium as $I^{\text{in}} $ and $ I^{\text{out}} $, respectively. Then, from  \cite{hodgson} and \cite{Carroll}, $I^{\text{in}} $ is expressed as 
     \begin{equation}
     \label{ap_I_in}
     	I^{\text{in}}  = \frac{I_\text{s}\hat{\delta}(g_0l + \ln \sqrt{\hat{\delta}} )}{1-\hat{\delta}},
     \end{equation}
     where $ \hat{\delta} $ is the total energy loss between the gain medium and the right side retroreflector  $ \text{R}_2 $ satisfying   $ I^{\text{in}} = I^{\text{out}}\hat{\delta} $ (For simplicity, we consider the case that there is no energy loss between the gain medium and the left side retroreflector $ \text{R}_1 $). Beside, $g_0 $ is the unsaturated gain coefficient, which is given as\cite{laserbook1}
 	\begin{equation}
	\label{ap_g_0}
		g_0=\frac{\eta P_\text{in}}{I_\text{s}lS_0}.
	\end{equation}  
      Moreover, since the laser intensity is stable, $ \hat{\delta} $ satisfies\cite{Carroll}
     \begin{equation}
     \label{ap_bal}
     	G(I^{\text{in}})\hat{\delta} = 1.
     \end{equation}
     Then, combining \eqref{ap_I_in} and \eqref{ap_bal} and replacing $I^{\text{in}}$  with $ I^{\text{T}, \text{in}}_k(n) $, we obtain
     \begin{equation}
     	   \label{ap_G_calculate}
     	I^{\text{T}, \text{in}}_k(n) = \frac{\left(g_0l - \ln \sqrt{G\left(I^{\text{T}, \text{in}}_k(n)\right)} \right)I_\text{s}}{G\left(I^{\text{T}, \text{in}}_k(n)\right)-1}.
     \end{equation}

	Finally, \eqref{G_calculate} is obtained by substituing \eqref{ap_g_0} into \eqref{ap_G_calculate}. 
       From \eqref{ap_G_calculate}, it is easy to see that $ G\big(I^{\text{T},\text{in}}_k(n)\big) >1 $ since input intensity 
       $ I^{\text{T},\text{in}}_k(n) $ is larger than 0. Besides, input intensity  $ I^{\text{T},\text{in}}_k(n) $ monotonically decreases as power gain $ G\big(I^{\text{T},\text{in}}_k(n)\big) $ increases\cite{Carroll}.
     Therefore, $ G\big(I^{\text{T},\text{in}}_k(n)\big) $ can be uniquely derived from \eqref{G_calculate} for the given $ I^{\text{T},\text{in}}_k(n) $.
     Moreover, power gain function $ G(\cdot) $ has the following properties.
     	\begin{Lemma}
     \label{le_I_convex} From the proof of \eqref{G_calculate}, it follows:
       \begin{enumerate}
     	  \item Power gain $G\big(I^{\text{T}, \text{in}}_k(n)\big)$  monotonically decreases as input intensity $ I^{\text{T}, \text{in}}_k(n) $  increases, and it satisfies $ 1 < G\big(I^{\text{T}, \text{in}}_k(n)\big) < e^{\frac{2\eta P_\text{in}}{I_\text{s}S_0}} $.
     	  \item  Output intensity  $ I^{\text{T}, \text{out}}_k(n) $  monotonically increases  as input intensity $ I^{\text{T}, \text{in}}_k(n) $ increases.
        \end{enumerate}	  	
	 \end{Lemma}
     \begin{IEEEproof}
     The decreasing property of $G\big(I^{\text{T}, \text{in}}_k(n)\big)$ with respect to $ I^{\text{T}, \text{in}}_k(n) $ can be easily derived from \eqref{ap_G_calculate}. To prove other properties of  $G\big(I^{\text{T}, \text{in}}_k(n)\big)$, 
      We denote $ G =  G\big(I^{\text{T}, \text{in}}_k(n)\big) $, $ I^{\text{in}} = I^{\text{T}, \text{in}}_k(n) $, and $ I^{\text{out}} = I^{\text{T}, \text{out}}_k(n) $ to simplify the notations in this proof. From \eqref{ap_G_calculate}, we have $ \frac{I_\text{s}(g_0l - \ln \sqrt{G})}{G-1}>0 $ since $ I^{\text{in}} $ is always larger than 0, which implies $ 1<G<e^{2g_0l} $. Combining with \eqref{ap_g_0}, we have $  1<G<e^{\frac{2\eta P_\text{in}}{I_\text{S}S_0}}  $. Moreover, \eqref{ap_G_calculate} can be equivalently written as  
		\begin{equation}
			I^{\text{in}}G - I^{\text{in}}  = I_\text{s}g_0l - \frac{I_\text{s}}{2}\ln G.
			\label{ap_simple_g}
		\end{equation}
 		Considering the case that $ I_\text{s} $, $ g_0 $, and $ l  $ are all fixed, and simply differentiating both sides of equation \eqref{ap_simple_g}  with respect to $ I^{\text{in}} $, we have
 		\begin{equation}
 			\frac{\mathrm{d}G}{\mathrm{d} I^{\text{in}}} = \frac{1-G}{ I^{\text{in}} + \frac{I_\text{s}}{2G}}, \label{ap_diff_G}
 		\end{equation}
		implying that $\frac{\mathrm{d}G}{\mathrm{d} I^{\text{in}}}$ is always less than $ 0 $ with $ 1< G <  e^{2g_0l}$. Therefore, $ G $ decreases with respect to $ I^{\text{in}} $ and part 1) is proved.
		
			 Similar to the above analysis, we complete the proof of part 2) for this proposition as follows. From \eqref{G_defi}, we have $ I^{\text{out}} = GI^{\text{in}}  $. Then, together with \eqref{ap_diff_G},  the first-order derivative of $ I^{\text{out}} $ with respect to $ I^{\text{in}} $ is given as
			\begin{equation}
				\frac{\mathrm{d}I^{\text{out}}}{\mathrm{d} I^{\text{in}}} = 1 + \frac{I_\text{s}(1-\frac{1}{G})}{ 2I^{\text{in}} + \frac{I_\text{s}}{G}},
			\end{equation} 
			which implies $ \frac{\mathrm{d}I^{\text{out}}}{\mathrm{d} I^{\text{in}}} >0 $ due to $  1<G<e^{\frac{2\eta P_\text{in}}{I_\text{S}S_0}}  $ and $ I^{\text{in}} >0 $. Therefore, we have completed this proof. 	 	
     \end{IEEEproof}
  
\section{Proof of Proposition \ref{prop_h}}
\label{ap_prop_h}

As depicted in Fig. \ref{figure_system}, at frame $ k$, we define $ I_k^{\text{R}, \text{in}}(n)  $ and $ I_k^{\text{R}, \text{out}}(n)  $ as the input and output beam intensities of the gain medium at the receiver, respectively. Then, from \eqref{G_defi}, we have
\begin{equation}
   I_k^{\text{R}, \text{out}}(n) = G \left(I_k^{\text{R}, \text{in}}(n)\right)I_k^{\text{R}, \text{in}}(n).	
   \label{ap_R_out}
\end{equation}
Moreover, since $ I_k^{\text{R}, \text{in}}(n) $ is the intensity of the resonant beam transmitted from the transmitter to the receiver and split by an optical splitter with power ratio $ 1-\alpha $, it satisfies 
\begin{equation}
	I_k^{\text{R}, \text{in}}(n) = \frac{(1-\alpha) \delta x_k^2(n)}{S_0},
	\label{ap_Rin}
\end{equation}
where $ S_0 $ is the cross-sectional area of the gain medium shown in Fig. \ref{gain_part1}. From Lemma \ref{le_I_convex}, it can be easily obtained that $ I_k^{\text{R}, \text{out}}(n) $ monotonically increases as  $ x_k(n) $ increases. Then, after being amplified by the gain medium at the receiver, resonant beam with intensity $ I_k^{\text{R}, \text{out}}(n) $ is transmitted back to the transmitter for the next frame transmission. By considering the scenario that the link loss from the receiver to the transmitter is equal to that of the reverse link, $ I_{k+1}^{\text{T}, \text{in}}(n) $ and $I_{k+1}^{\text{T}, \text{out}}(n)  $ are given as
\begin{equation}
I_{k+1}^{\text{T}, \text{in}}(n) = \delta I_k^{\text{R}, \text{out}}(n), \label{ap_I_T_in}
\end{equation}
and 
\begin{equation}
	I_{k+1}^{\text{T}, \text{out}}(n) = G\left(I_{k+1}^{\text{T}, \text{in}}(n)\right)I_{k+1}^{\text{T}, \text{in}}(n),
	\label{ap_T_out}
\end{equation}
respectively, where \eqref{ap_T_out} is obtained by \eqref{G_defi}. Then, from \eqref{ap_I_T_in}, \eqref{ap_T_out}, and Lemma \ref{le_I_convex} of Appendix \ref{ap_prop_I},  it is easy to show that $ I_{k+1}^{\text{T}, \text{out}}(n) $ monotonically increases as  $ I_k^{\text{R}, \text{out}}(n) $ increases.
Since $ h(x_k(n)) $ represents the output power of the gain medium at the transmitter by \eqref{x_k}, it follows  $ h(x_k(n)) = I_{k+1}^{\text{T}, \text{out}}(n)S_0 $. Therefore, we can easily derive that $ h(x_k(n)) $  monotonically increases as  $ x_k(n) $ increases. Moreover, we obtain
    \begin{align}
    	h(x_k(n)) &= I_{k+1}^{\text{T}, \text{out}}(n)S_0 \notag \\
    	 &= G\left(I_{k+1}^{\text{T}, \text{in}}(n)\right)I_{k+1}^{\text{T}, \text{in}}(n)S_0 \label{ap_h_1} \\
    	&= G\left(I_k^{\text{R}, \text{out}}(n)\delta\right)I_k^{\text{R}, \text{out}}(n) S_0\delta  \label{ap_h_2}\\  
    	&= G\left(G \left(I_k^{\text{R}, \text{in}}(n)\right)I_k^{\text{R}, \text{in}}(n)\delta \right)G \left(I_k^{\text{R}, \text{in}}(n)\right)I_k^{\text{R}, \text{in}}(n) S_0\delta, \label{ap_h_3}
    \end{align}
    where \eqref{ap_h_1}, \eqref{ap_h_2}, and \eqref{ap_h_3} are obtained by \eqref{ap_T_out}, \eqref{ap_I_T_in}, and \eqref{ap_R_out}, respectively. Finally, \eqref{func_h} is obtained by replacing $ I_k^{\text{R}, \text{in}}(n)$ in \eqref{ap_h_3} with \eqref{ap_Rin}.

   \begin{Lemma}
   \label{le_h} From the proof of \eqref{func_h}, link gain function $ h(x_k(n)) $ has the following properties:
    \begin{enumerate}
   	   \item  $ h(x_k(n)) $  monotonically increases as  $ x_k(n) $ increases, and satisfies $ 0<x_{k}(n)$ and $\sqrt{h(x_{k}(n))}  \leq \sqrt{P_\text{t}}  $.
    	\item  $ \frac{\sqrt{h(x_k(n))} }{x_k(n)}  $ monotonically decreases as $ x_k(n) $ increases, and it satisfies $ 1\leq \frac{\sqrt{h(x_k(n))} }{x_k(n)} < e^{\frac{2\eta P_\text{in}}{I_\text{s}S_0}}\delta \sqrt{(1-\alpha)} $.
    \end{enumerate}
   \end{Lemma}
\begin{IEEEproof}
   The increasing property of $ h(x_k(n)) $ with respect to $ x_k(n) $ has been proved when deriving the expression for $h(\cdot)$.  From \eqref{x_k},  we have $ x_1(n) \leq\sqrt{P_\text{t}} $, with the equality holding if and only if $ m_1(n)=1 $. Since $ P_\text{t} $ is the stable power of the resonant beam at the transmitter, it follows $ h(\sqrt{P_\text{t}})= P_\text{t} $. Furthermore, the monotonicity of $ h(\cdot) $ implies $ h(x_1(n)) \leq P_\text{t} $, and consequently, it follows $ x_2(n)\leq \sqrt{P_\text{t}} $ by \eqref{x_k}. Similarly, we have $ 0<x_{k}(n)$ and $\sqrt{h(x_{k}(n))}  \leq \sqrt{P_\text{t}}  $, $ \forall k = 1,2,3\cdots$,   by mathematical induction.
    
   To prove part 2), by utilizing \eqref{func_h} and \eqref{ap_h_2}, we have
   \begin{equation}
   	 \frac{h(x_k(n))}{x_k^2(n)}=(1-\alpha)\delta^2 G\left(I_k^{\text{R}, \text{out}}(n)\delta \right)G\left(\frac{(1-\alpha)\delta x_k^2(n) }{S_0}\right).   \label{ap_h_4}
   	 \end{equation}
  Since $ I_k^{\text{R}, \text{out}}(n) $ monotonically increases  as  $ x_k(n) $ increases, both $ G\big(I_k^{\text{R}, \text{out}}(n)\delta \big) $  and $ G\left(\frac{(1-\alpha)\delta x_k^2(n) }{S_0}\right) $ monotonically decrease with respect to $ x_k(n) $ by Lemma \ref{le_I_convex} of Appendix \ref{ap_prop_I}.  Therefore, $ \frac{\sqrt{h(x_k(n))} }{x_k(n)}  $  monotonically decreases as $ x_k(n) $ increases.  Moreover, as $ 0< x_k(n) \leq \sqrt{P_\text{t}} $, it is easy to check $ 1 = \frac{h(\sqrt{P_\text{t}}) }{P_\text{t}} \leq \frac{h(x_k(n))}{x_k^2(n)} $.  Besides, $ \frac{h(x_k(n))}{x_k^2(n)} < (1-\alpha)\delta^2 e^{\frac{4\eta P_\text{in}}{I_\text{s}S_0}} $ is obtained by \eqref{ap_h_4} and Lemma \ref{le_I_convex}. Therefore, part 2) has been proved.

\end{IEEEproof}
    
\section{Proof of Proposition \ref{prop_wk} }
\label{ap_prop_wk}

From \eqref{m_k}, it is easy to see that $ s_k(n) $ is bounded over $ (0,1] $, due to $ m_k(n) \in (0,1] $ by \eqref{x_k}, $ w_k(n)\in (0,1] $, and the i.i.d. property of  $ s_k(n)$'s. For simplicity, we denote $ \mu_1(n) $ as the minimum value of $s_k(n) $. It is obvious to derive that the design scheme \eqref{m_k}-\eqref{w_k} holds for the case of $ k =1 $ if and only if  $0< A_n \leq \sqrt{P_\text{t}} $. For the case of $ k \geq 2 $, we first prove the ``if'' part by mathematical induction. 
   When $ k = 2 $, we have 
   \begin{align}
   	\sqrt{h(x_{1}(n))} &=  	\sqrt{h\big(\sqrt{P_\text{t}}w_1(n)s_1(n)\big)} \label{p_h_1} \\ 
   	   & \geq \sqrt{h\big(A_n\mu_1(n) \big)}  \label{p_h_2}  \\  
   	   & \geq \sqrt{h\big(\hat{A}_n\big)} = A_n,     \label{p_h_3}
   \end{align}
   where \eqref{p_h_1} is obtained by \eqref{x_k} and \eqref{m_k}, \eqref{p_h_2} is derived by \eqref{w_k} when $ k =1 $, and \eqref{p_h_3} is obtained due to the monotonically increasing property of link gain function $ h(x_{k-1}(n)) $ given in Lemma \ref{le_h} of Appendix \ref{ap_prop_h}. Therefore, there  exists $ w_2(n) \in (0, 1] $ such that $ \sqrt{h(x_1(n))}w_2(n) = A_n $.  Then, suppose that \eqref{m_k}-\eqref{w_k} is true   for $ k = k_0 $ $(k_0\geq2)$.  Similar to \eqref{p_h_1} -- \eqref{p_h_3}, we have 
   \begin{align}
   	\sqrt{h(x_{k_0}(n))} &= \sqrt{h\big(\sqrt{h(x_{k_0-1}(n))}w_{k_0}(n)s_{k_0}(n)\big)} \label{p_h_k1} \\ 
   	&= \sqrt{h\big(A_ns_{k_0}(n) \big)} \quad  \label{p_h_k2}  \\
   	&\geq \sqrt{h\big(A_n\mu_1(n) \big)} \geq A_n, \label{p_h_k3}
   \end{align}
   where \eqref{p_h_k2} is obtained by the induction hypothesis.  
  It means that there  exists $ w_{k_0+1}(n) \in (0, 1] $ such that $ \sqrt{h(x_{k_0}(n))}w_{k_0+1}(n) = A_n $  is true. Thus, the design scheme \eqref{m_k}-\eqref{w_k} holds for $ k = k_0 +1 $, and the proof of the induction step is complete.
   
   We then prove the ``only if'' part by contradiction. Suppose that  $ \mu_1(n) < \frac{\hat{A}_n}{A_n} $ is true and the design scheme \eqref{m_k}-\eqref{w_k}  holds for all $ k \geq 2 $. Together with \eqref{p_h_k1} -- \eqref{p_h_k2},  we obtain
   \begin{align}
   	       \min\limits_{s_{k_0}(n)}\sqrt{h(x_{k_0}(n))} &= \sqrt{h\big(A_n\mu_1(n) \big)} \\
   	       &< \sqrt{h(\hat{A}_n)} = A_n.
   \end{align}
   Since $ w_{k_0+1} \in(0, 1] $,  the design scheme \eqref{m_k}-\eqref{w_k} does not hold for $ k = k_0 +1 $, which brings a contradiction.   Therefore, \eqref{w_k} is true for $ k\geq 2 $ only if $ \mu_1(n) \geq \frac{\hat{A}_n}{A_n} $ holds.  Then, the whole proof is completed.

\section{Proof of Lemma \ref{le_Pt} }

\begin{figure}[t]
 	\centering
 	\includegraphics[width=4.5in]{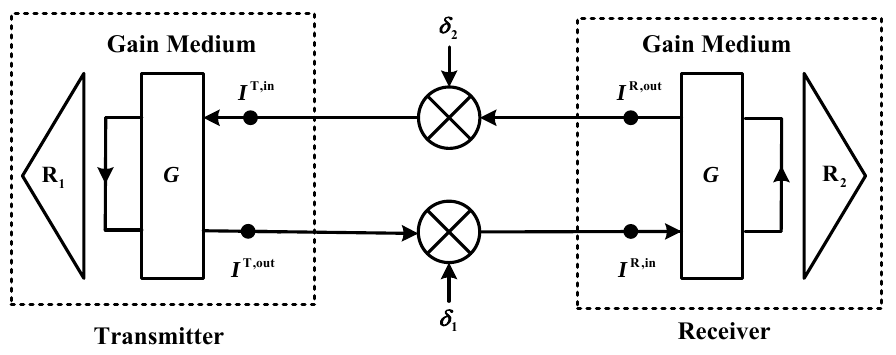}
 	\caption{Schematic diagram of the gain process of the energy flow process when the resonant beam is stable.} 
 	\label{ap_fig_energy_flow}
    \end{figure}

    To prove this lemma, we define the total link loss from the transmitter to the receiver and the total link loss from the receiver to the transmitter as $ \delta_1 $ and $ \delta_2 $, respectively. Fig. \ref{ap_fig_energy_flow} shows the energy flow process when the resonant beam is stable, where we denote $ I^{\text{T}, \text{in}} $ and $ I^{\text{T}, \text{out}} $ as the stable input and output beam intensity of the gain medium at the transmitter, respectively, and denote $ I^{\text{R}, \text{in}} $ and $ I^{\text{R}, \text{out}} $ as the stable input and output beam intensity of the gain medium at the receiver, respectively.  Using the definition of link gain function $h(\cdot)$ in \eqref{func_h}, we can derive that $\delta_1 = (1-\alpha)\delta$ and $\delta_2 = \delta$.   Moreover,  from the proof of Lemma \ref{le_h} in Appendix \ref{ap_prop_h}, it is easy to show that $ P_\text{t} $ monotonically increases with respect to  $ \delta_1 $ and $ \delta_2 $, respectively. 
    \label{ap_le_Pt}
   Denote the upper and lower bounds of  $ P_\text{t} $ as $ P_{\text{up}} $ and $ P_{\text{low}} $, respectively. In order to derive $ P_{\text{up}}  $, we set $ \delta_1 = \delta $ since $ 0 <\alpha <1 $.  Then, as shown in Fig. \ref{ap_fig_energy_flow},  given that the transmitter and receiver gain mediums have the same parameters, and $\delta_1=\delta_2=\delta$,  we have\cite{Carroll}  
   \begin{equation}
   	G(I^{\text{T}, \text{in}}) = G(I^{\text{R}, \text{in}})= \frac{1}{\delta}, \label{ap_G_stable}
   \end{equation} and 
   \begin{equation}
   	I^{\text{T}, \text{in}} =  I^{\text{R}, \text{in}} = P_{\text{up}}\delta/S_0. \label{ap_P_t_in}
   \end{equation}
   Then, by replacing $ G\big(I^{\text{T}, \text{in}}_k(n)\big) $ and $ I^{\text{T}, \text{in}}_k(n) $ in \eqref{G_calculate} with  \eqref{ap_G_stable} and \eqref{ap_P_t_in}, respectively, $ P_{\text{up}} $ is obtained as 
   \begin{equation}
   P_{\text{up}}=	\frac{2\eta P_\text{in} + I_\text{s}S_0\ln \delta  }{2(1-\delta )}. 
   \end{equation}
    Similarly, $ P_{\text{low}} $ is obtained by setting  $ \delta_1 = \delta_2 = (1-\alpha)\delta  $. Therefore, this lemma is proved.

%


    \end{document}